\documentclass{emulateapj}
\usepackage{graphicx}
\usepackage[flushleft]{threeparttable}

\slugcomment{Accepted for publication in ApJ}
\shorttitle{A molecular line scan in the HDF--N}
\shortauthors{Decarli et al.}

\def\Lsun{L$_\odot$}
\def\Msun{M$_\odot$}

\def\Oiii{[O\,{\sc iii}]}

\def\Ci{[C\,{\sc i}]}

\def\Cii{[C\,{\sc ii}]}
\def\Ci{[C\,{\sc i}]}

\def\kms{km\,s$^{-1}$}

\def\jykms{Jy~km\,s$^{-1}$}
\def\Kkmspc{K~km\,s$^{-1}$\,pc$^2$}

\def\lsim{\mathrel{\rlap{\lower 3pt \hbox{$\sim$}} \raise 2.0pt \hbox{$<$}}}
\def\gsim{\mathrel{\rlap{\lower 3pt \hbox{$\sim$}} \raise 2.0pt \hbox{$>$}}}

\begin{document}

\title{
A molecular line scan in the Hubble Deep Field North
}

\author{
Decarli R.\altaffilmark{1}, 
Walter F.\altaffilmark{1},
Carilli C.\altaffilmark{2},
Riechers D.\altaffilmark{3},
Cox P.\altaffilmark{4},
Neri R.\altaffilmark{4},
Aravena M.\altaffilmark{5,6},
Bell E.\altaffilmark{7},
Bertoldi F.\altaffilmark{8},
Colombo D.\altaffilmark{1},
Da Cunha E.\altaffilmark{1},
Daddi E.\altaffilmark{9},
Dickinson M.\altaffilmark{10},
Downes D.\altaffilmark{4},
Ellis R.\altaffilmark{11},
Lentati L.\altaffilmark{12},
Maiolino R.\altaffilmark{12},
Menten K.M.\altaffilmark{13},
Rix H.--W.\altaffilmark{1},
Sargent M.\altaffilmark{9},
Stark D.\altaffilmark{14},
Weiner B.\altaffilmark{14},
Weiss A.\altaffilmark{13}
}
\altaffiltext{1}{Max-Planck Institut f\"{u}r Astronomie, K\"{o}nigstuhl 17, D-69117, Heidelberg, Germany. E-mail: {\sf decarli@mpia.de}}
\altaffiltext{2}{NRAO, Pete V.\,Domenici Array Science Center, P.O.\, Box O, Socorro, NM, 87801, USA}
\altaffiltext{3}{Cornell University, 220 Space Sciences Building, Ithaca, NY 14853, USA}
\altaffiltext{4}{IRAM, 300 rue de la piscine, F-38406 Saint-Martin d'H\`eres, France}
\altaffiltext{5}{European Southern Observatory, Alonso de Cordova 3107, Casilla 19001, Vitacura Santiago, Chile}
\altaffiltext{6}{N\'{u}cleo de Astronom\'{\i}a, Facultad de Ingenier\'{\i}a, Universidad Diego Portales, Av. Ej\'{e}rcito 441, Santiago, Chile}
\altaffiltext{7}{Department of Astronomy, University of Michigan, 500 Church St., Ann Arbor, MI 48109, USA}
\altaffiltext{8}{Argelander Institute for Astronomy, University of Bonn, Auf dem H\"{u}gel 71, 53121 Bonn, Germany}
\altaffiltext{9}{Laboratoire AIM, CEA/DSM-CNRS-Universite Paris Diderot, Irfu/Service d'Astrophysique, CEA Saclay, Orme des Merisiers, 91191 Gif-sur-Yvette cedex, France}
\altaffiltext{10}{National Optical Astronomy Observatory, 950 North Cherry Avenue, Tucson, Arizona 85719, USA}
\altaffiltext{11}{Astronomy Department, California Institute of Technology, MC105-24, Pasadena, California 91125, USA}
\altaffiltext{12}{Cavendish Laboratory, University of Cambridge, 19 J J Thomson Avenue, Cambridge CB3 0HE, UK}
\altaffiltext{13}{Max-Planck-Institut f\"ur Radioastronomie, Auf dem H\"ugel 69, 53121 Bonn, Germany}
\altaffiltext{14}{Steward Observatory, University of Arizona, 933 N. Cherry St., Tucson, AZ  85721, USA}

\begin{abstract}
We present a molecular line scan in the Hubble Deep Field North (HDF--N) that covers the entire 3\,mm window (79--115 GHz) using the IRAM Plateau de Bure Interferometer. Our CO redshift coverage spans $z\lsim0.45$, $1\lsim z\lsim 1.9$ and all $z\gsim 2$. We reach a CO detection limit that is deep enough to detect essentially all $z>1$ CO lines reported in the literature so far. We have developed and applied different line searching algorithms, resulting in the discovery of 17 line candidates. We estimate that the rate of false positive line detections is $\sim2/17$. We identify optical/NIR counterparts from the deep ancillary database of the HDF--N for seven of these candidates and investigate their available SEDs. Two secure CO detections in our scan are identified with star-forming galaxies at $z=1.784$ and at $z=2.047$. These galaxies have colors consistent with the `BzK' color selection and they show relatively bright CO emission compared with galaxies of similar dust continuum luminosity. We also detect two spectral lines in the submillimeter galaxy HDF850.1 at $z=5.183$. We consider an additional 9 line candidates as high quality. Our observations also provide a deep 3\,mm continuum map (1-$\sigma$ noise level = 8.6 $\mu$Jy\,beam$^{-1}$). Via a stacking approach, we find that optical/MIR bright galaxies contribute only to $<50$\% of the SFR density at $1<z<3$, unless high dust temperatures are invoked. The present study represents a first, fundamental step towards an unbiased census of molecular gas in `normal' galaxies at high--$z$, a crucial goal of extragalactic astronomy in the ALMA era.
\end{abstract}

\section{Introduction}

Studies of the molecular medium in distant galaxies provide key diagnostics about the evolutionary state of galaxies in the high--redshift universe. To date, $\sim$200 high--redshift galaxies have been detected in CO emission, the main tracer for molecular gas \citep[e.g., review by][]{carilli13}. Essentially all these detections were obtained by targeted observations of galaxies that had been pre--selected through their star forming properties (e.g. UV or FIR emission).

\begin{figure}
\includegraphics[width=0.99\columnwidth]{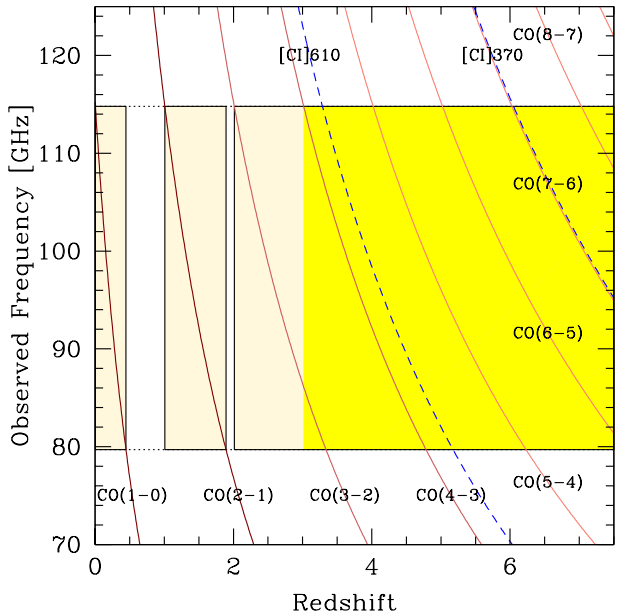}\\
\caption{CO redshift coverage of our molecular line scan (from 79.7--114.8\,GHz). We sample CO emission at $0.00412<z<0.4464$, $1.008<z<1.893$, and at any $z>2.012$, as highlighted by the shaded area. Darker shading marks the redshift range ($z>3.016$) for which multiple lines (CO or \Ci{}) are simultanously covered.}
\label{fig_z_range}
\end{figure}

Advances in millimeter technology now allow spectral-line scans across full atmospheric windows. Scans over a broad frequency range now allow meaningful attempts at redshift determinations in heavily-obscured, high-$z$ galaxies, through the detection of molecular gas tracers. In turn, they enable studies of the molecular gas emission over large cosmic volumes. Such studies have been out of reach until very recently, because of limitations both in sensitivity and instantaneous frequency coverage of existing millimeter facilities. Here we present the first blind molecular line scan of an optical deep field, i.e. a region in the sky for which superb multi--wavelength data is available. Our target of choice is a region in the Hubble Deep Field North \citep[HDF--N,][]{williams96}, one of the best studied optical/NIR deep fields. We have used the IRAM Plateau de Bure Interferometer (PdBI) to scan the 3\,mm atmospheric band, to search for CO transitions. This band is particularly well suited for CO surveys of high-$z$ galaxies, as it targets the lowest--J CO transitions, which are collisionally excited by H$_2$ at low temperatures. As shown in Fig.~\ref{fig_z_range} and Tab.~\ref{tab_z_range}, the CO and \Ci{} redshift ranges probed in a 3\,mm scan are $z\lsim0.5$, and $1.0\lsim z\lsim1.9$, and $z\gsim2$. At $z\gsim3$ two or more CO transitions are covered by such a scan, thus allowing unambiguous redshift identifications. In \citet{walter12}, we used the dataset presented here to search for CO emission associated with HDF\,850.1, the brightest sub--millimeter galaxy in the HDF--N \citep[HDF--N,][]{williams96, hughes98}. This search allowed us to unambiguously establish the redshift of this luminous galaxy \citep[$z=5.183$,][]{walter12}. A similar `blind line search' approach was presented in \citet{weiss13}, who performed a scan of the 3\,mm window with ALMA to obtain redshifts for 26 strongly--lensed sub-mm galaxies (SMGs). This search resulted in the discovery of 44 lines, and in the identification of 20 unambiguous redshifts. The ALMA observations presented in \citet{weiss13} have a typical rms noise level of 2 mJy\,beam$^{-1}$ per 50-60 \kms{} channel, i.e., a 4-$\sigma$ line luminosity limit of $\sim5\times10^{10}$ \Kkmspc{} (assuming a line width of 300 \kms{}). This depth is insufficient to probe any CO emission arising from main sequence galaxies in the targeted fields. 

\begin{figure}
\includegraphics[width=0.99\columnwidth]{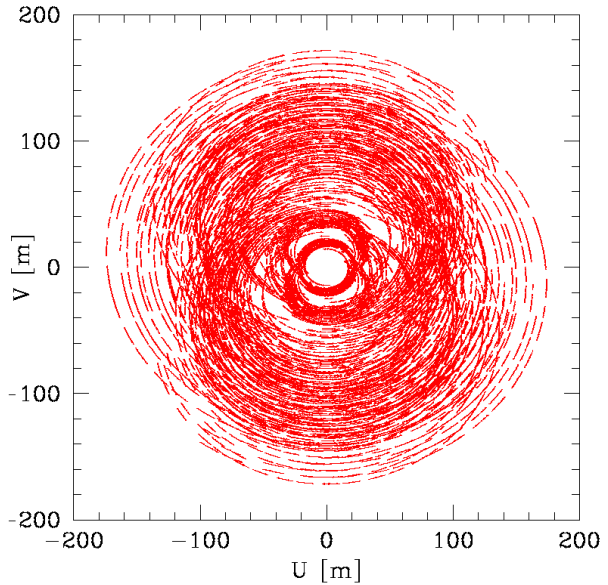}\\
\caption{The $(U,V)$ coverage of our observations. A total of 43 tracks and $132448$ visibilities (110 hr on source, 6-antenna equivalent) spread over 10 different frequency setups were used to produce the final data cubes.}
\label{fig_uvcover}
\end{figure}
\begin{table}
\caption{\rm Lines and corresponding redshift ranges covered in our molecular line scan (79.7--114.8\,GHz). Comoving volume and volume--weighted average redshifts are computed by accounting for the changing size of the primary beam as a function of observed frequency.} \label{tab_z_range}
\begin{center}
\begin{tabular}{cccccc}
\hline
Transition  & $\nu_0$ & $z_{\rm min}$ & $z_{\rm max}$ & $\langle z \rangle$ & Volume \\
            & [GHz]   &     &        &       & [Mpc$^3$] \\
    (1)     & (2)     & (3) & (4)    & (5)   & (6) \\
\hline
CO(1-0) 	& 115.271 & 0.00412 & 0.446 & 0.338 &	92 \\
CO(2-1) 	& 230.538 & 1.008   & 1.893 & 1.525 & 1442 \\
CO(3-2) 	& 345.796 & 2.012   & 3.339 & 2.751 & 2437 \\
CO(4-3) 	& 461.041 & 3.016   & 4.785 & 3.981 & 2966 \\
CO(5-4) 	& 576.268 & 4.020   & 6.231 & 5.213 & 3249 \\
CO(6-5) 	& 691.473 & 5.023   & 7.676 & 6.445 & 3403 \\
CO(7-6) 	& 806.652 & 6.027   & 9.122 & 7.679 & 3484 \\
\hline
\Ci{}$_{1-0}$   & 492.161 & 3.287   & 5.175 & 4.313 & 1550 \\
\Ci{}$_{2-1}$   & 809.342 & 6.050   & 9.155 & 7.707 & 3508 \\
\hline
\end{tabular}
\end{center}
\end{table}

The molecular line scan (`blind CO search') presented here is the first one that is sensitive enough to probe the molecular content beyond the `tip of the iceberg' of luminous SMGs. This paper describes the scan and first observational results. In \S\ref{sec_observations} we describe our observational setup. \S\ref{sec_method} presents the line--searching techniques adopted in our analysis. In \S\ref{sec_results} we describe our results from the line search, the properties of our 3\,mm--selected line candidates, and we investigate the CO properties of optically/NIR--selected galaxies. In \S\ref{sec_models} we compare our results with empirical predictions and \S\ref{sec_cont} presents our analysis of the 3\,mm continuum emission. Our conclusions are summarized in \S\ref{sec_conclusions}. Implications of our results regarding the cosmic abundance of molecular gas at high $z$ are presented in a companion paper (Walter et al.\ 2013; hereafter, W13).

Throughout the paper we will assume a standard cosmology with $H_0=70$ km s$^{-1}$ Mpc$^{-1}$, $\Omega_{\rm m}=0.3$ and $\Omega_{\Lambda}=0.7$.

\section{Observations}\label{sec_observations}

\begin{figure}
\includegraphics[width=0.99\columnwidth]{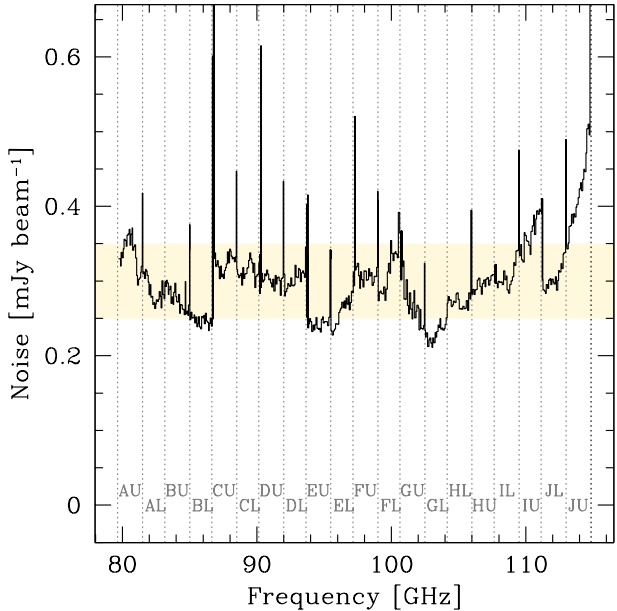}\\
\caption{Noise per 90 \kms{} channel in our molecular line scan. The different frequency setups of our observations are labeled. Spikes in the noise properties are observed at the edges of various frequency setups. The shaded box highlights the $(0.25-0.35)$ mJy\,beam$^{-1}$ range, our typical sensitivity limit.}
\label{fig_noise_scan}
\end{figure}

\begin{figure}
\includegraphics[width=0.49\textwidth]{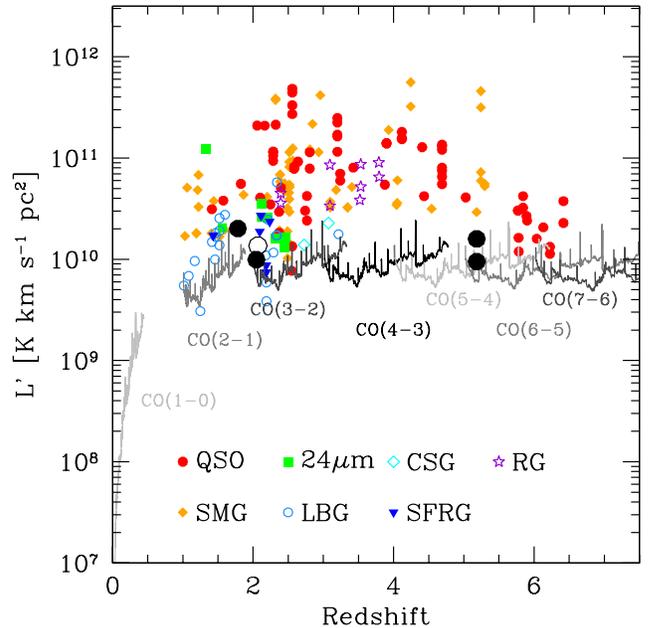}\\
\caption{The sensitivity of our 3\,mm scan in terms of line luminosity, for various CO transitions. The luminosity sensitivity is computed by assuming a line width of 300 \kms{}, a 3.5-$\sigma$ detection (see \S\ref{sec_observations}) and an average primary beam attenuation of $0.72$. The big filled circles mark the two CO detections from HDF850.1 at $z=5.183$ obtained in our molecular line scan, the CO(2--1) detection in ID.03 (the brightest line in our sample), and the CO(3--2) detection of ID.19, the only CO line in our survey for which a consistent spectroscopic redshift from the optical/NIR counterpart is available. The big empty circle marks the CO detection of ID.18, the only other line detected at $>5$--$\sigma$, for which we assume here a redshift $z=2.070$. Our observations are sensitive to CO line luminosities $(4-8) \times 10^{9}$ \Kkmspc{} at any $z>1$. For the sake of comparison, we overplot all $z>1$ CO detections to date from 3\,mm observations, from the compilation by \citet{carilli13} (here we have not applied de--magnification corrections). Colors and shapes of symbols mark different source types (quasars, sub-mm galaxies, 24-$\mu$m selected galaxies, Lyman-break galaxies, color-selected galaxies, star-forming radio galaxies, radio galaxies). With our achieved sensitivity, we would be able to detect essentially all previously detected CO emitters at high redshift.}
\label{fig_scan_lum}
\end{figure}

\subsection{The 3\,mm scan}
We have obtained a continuous scan of the 3\,mm window performed at the IRAM Plateau de Bure Interferometer (PdBI, pointing center: RA=12:36:50.300, Dec=+62:12:25.00, J2000.0)\footnote{This particular region within the HDF--N was chosen to include the bright sub-mm galaxy HDF850.1, see \citet{walter12}.}. The primary beam of PdBI can be described by a Gaussian profile with full width at half maximum (FWHM) = $47.3'' \times (100/\nu)$, where $\nu$ is the observing frequency in GHz. At the central frequency of our scan ($\nu=97.25$ GHz), the primary beam is $48.6''$ in diameter. Observations were obtained between December 06, 2010 and November 19, 2012, split into 43 tracks (projects \verb|U09E|, \verb|V0B6|, \verb|V--3|). We covered the entire 3\,mm window (79.696 GHz -- 114.798 GHz) with 10 frequency settings (labeled with capital letters from A to J with increasing frequency) using the PdBI WideX correlator. Observations were obtained in C configuration, in most cases using the full 6--antenna array, resulting in baselines from $14.5$ m to 176 m (see Fig.~\ref{fig_uvcover}). Flux calibration was achieved by observing various calibrators (in most cases: MWC349, B0923+392, 3C84, 3C273, 3C345, B0234+285, B2200+420, B1749+096). The quasar B1300+580 was used as a phase and amplitude calibrator. Data have been processed using the most recent version of the \textsf{GILDAS} software. The receiver operated in the lower sideband at frequencies below 104.2 GHz (setups A--G) and in the upper sideband in the remaining part of the scan (setups H--J). WideX has four different units, two observing the two polarization modes on the low-frequency side of the tuning frequency, and two observing the high-frequency side. We therefore extracted two cubes (at `U' and `L' with respect to the tuned frequency) for each frequency setup. We resampled the cubes in 90 \kms{} channels, and imaged them using the \textsf{GILDAS} suite \textsf{mapping}. Natural weighting was adopted. 

The typical beam size in our molecular line scan is $3''\times2.7''$ ($\sim$25 kpc at $z$=2, and roughly constant at any $z\gsim1$). The final cubes include $132,448$ visibilities in total, corresponding to $\approx 110$ hr on source (6--antenna equivalent). The final cubes have a typical rms of $0.3$ mJy beam$^{-1}$ in 90 \kms{} channels and we show the actual frequency dependence of the noise in Fig.~\ref{fig_noise_scan}. We can convert this limit into a CO luminosity sensitivity, by assuming a typical line width of 300 \kms{}, and an average primary beam correction of 0.72, and by requiring a 3.5-$\sigma$ line detection. The resulting CO luminosity limit as a function of redshift is shown in Fig.~\ref{fig_scan_lum}. At $z>1$, our $3.5$-$\sigma$ CO line detection threshold is $L'_{\rm lim}=(4-8)\times 10^{9}$ K\,km\,s$^{-1}$\,pc$^2$, comparable with the faintest CO detections reported so far from 3\,mm observations at these redshifts \citep{carilli13}.

\begin{figure*}
\includegraphics[width=0.49\textwidth]{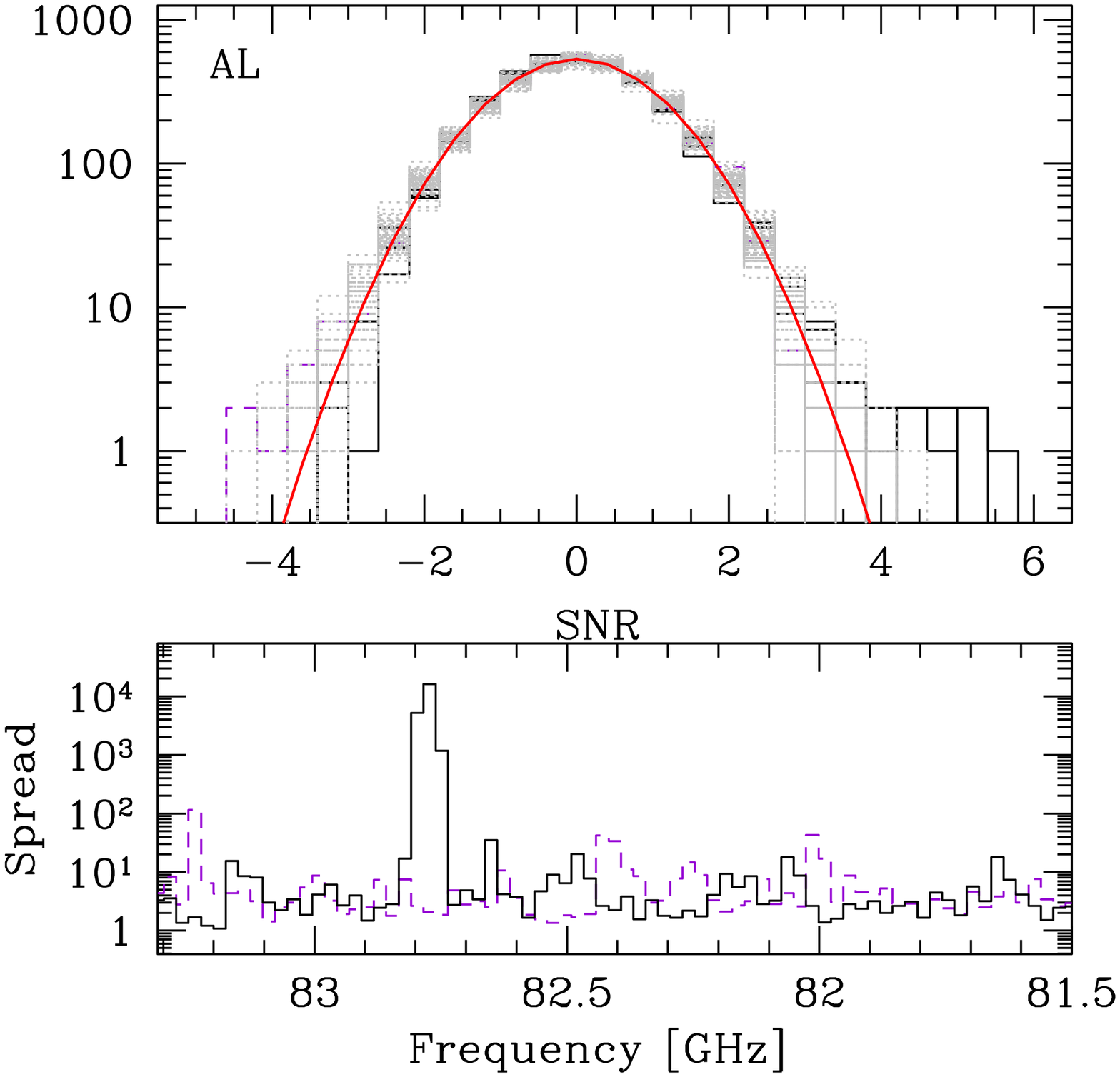}
\includegraphics[width=0.49\textwidth]{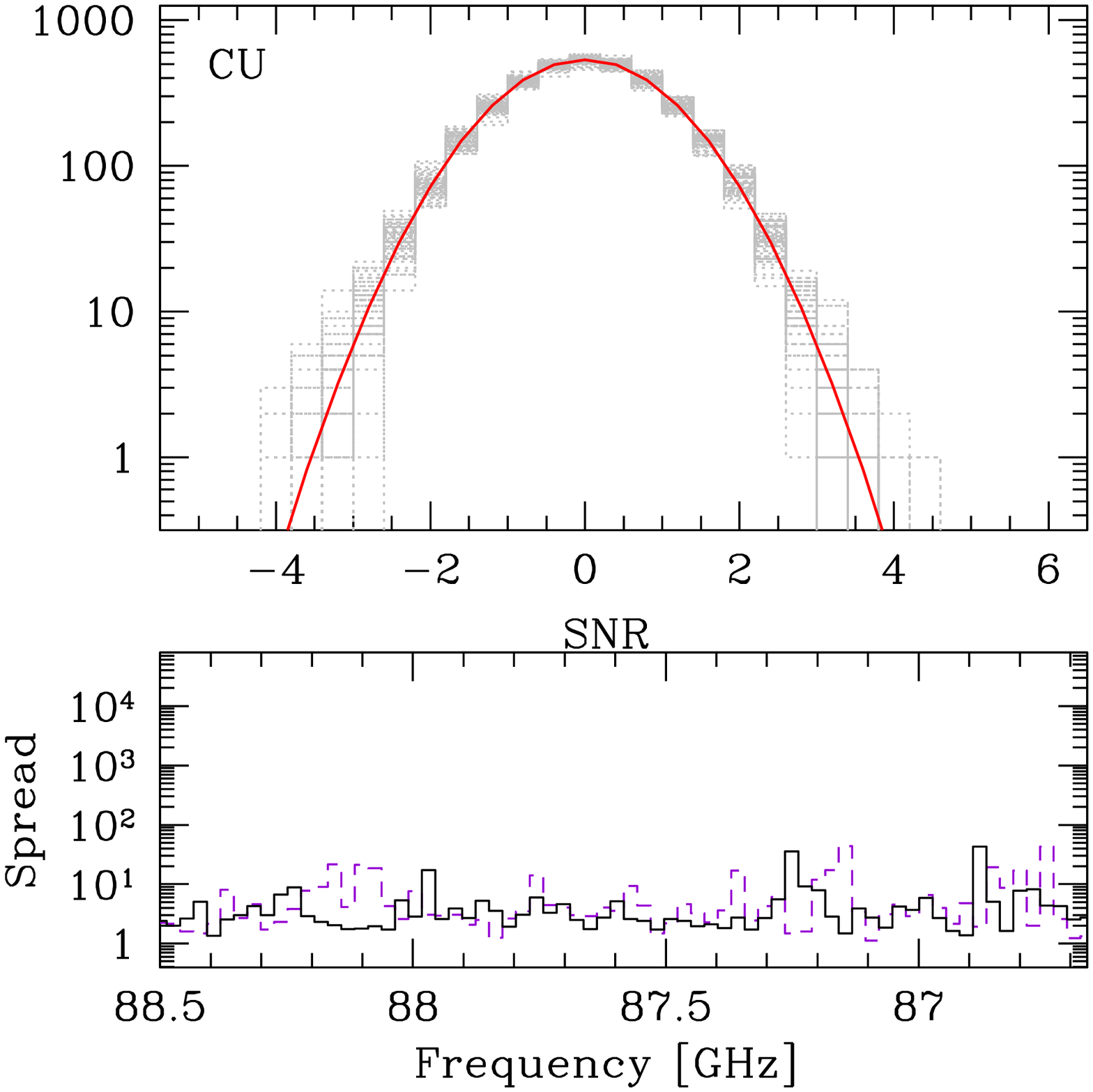}\\
\caption{Our {\em spread} analysis to detect molecular gas emission in the data cubes is applied to two example data cubes, AL and CU. In the top panel we plot, for each 90 \kms{} wide channel, the distribution of pixel values within $\sqrt{2}$ times the primary beam radius (gray, dotted histograms). Distributions are normalized using the rms for each channel map, and are compared to a gaussian with $\sigma$=1 and an area matching the number of pixels used in this analysis (red solid line). Significant positive (negative) excess is highlighted with black, solid (dashed purple) histograms. In the bottom panel, we show the distribution of the {\em spread} values (see \S\ref{sec_spread}, Eq.~\ref{eq_spread}) as a function of frequency. In the left panel (setup AL), positive excess is clearly seen at $\sim82.8$ GHz. A few minor positive and negative excesses are also reported in the AL cube. No significant deviations from a purely Gaussian distribution is reported in the CU cube, implying no line detection.}
\label{fig_spread}
\end{figure*}

%

\begin{figure*}
\includegraphics[width=0.53\textwidth]{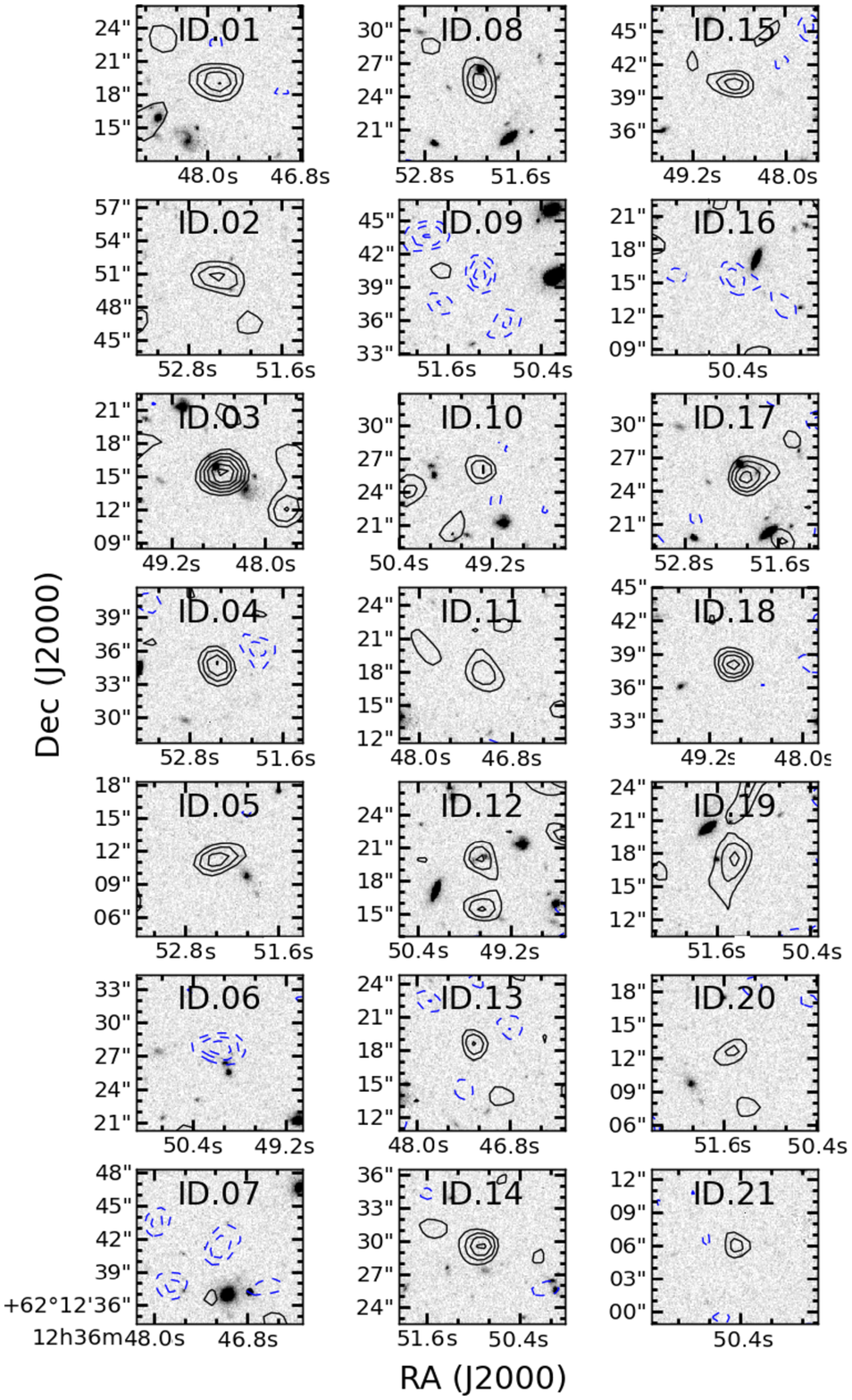}
\includegraphics[width=0.47\textwidth]{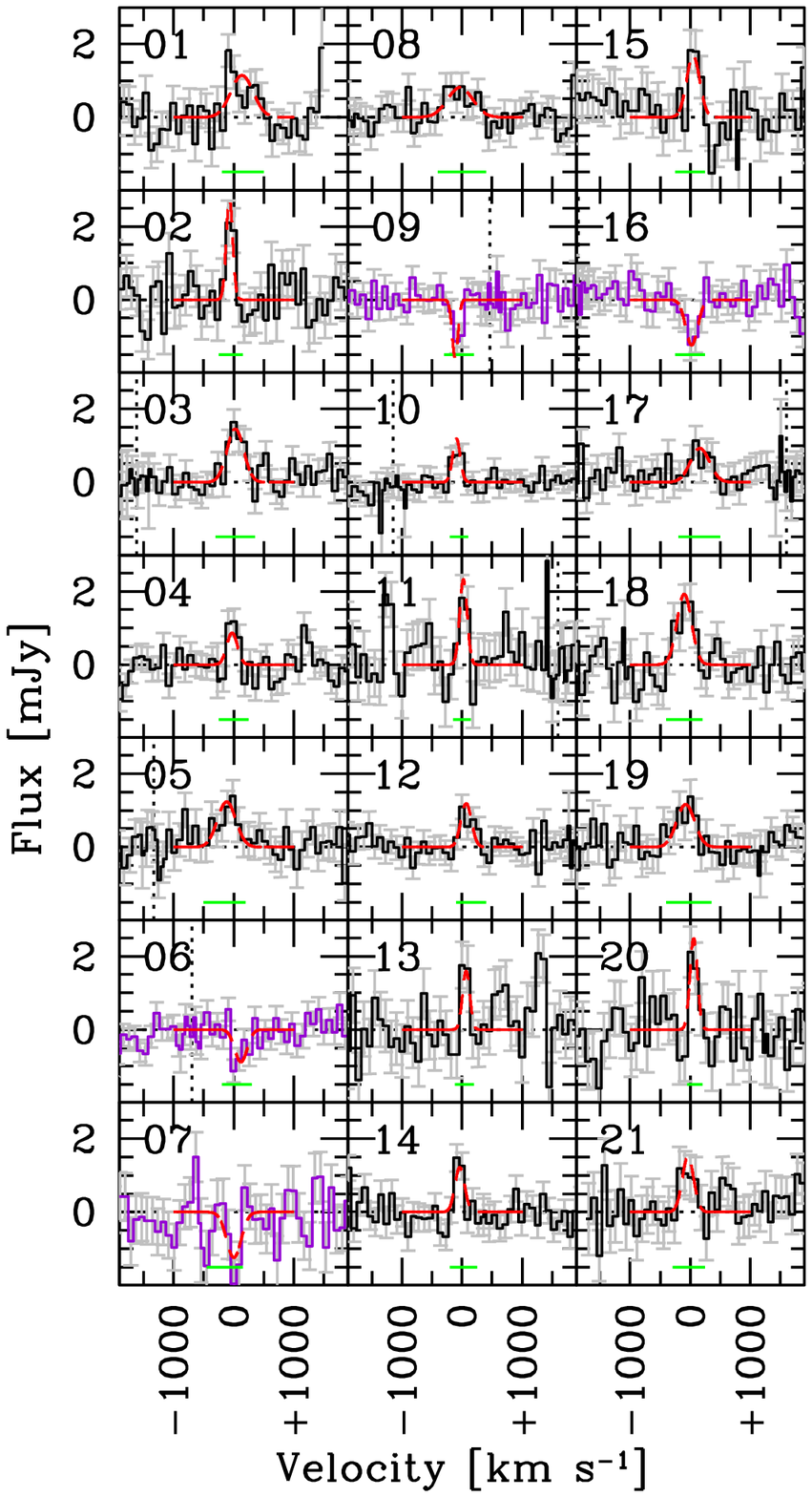}\\
\caption{Line maps ({\em left}) and spectra ({\em right}) of the line candidates revealed by our search (Tab.~\ref{tab_lines}). {\em Left:} Line maps are shown as contours (solid black / dashed blue contours show positive / negative [2, 3, 4, \ldots] $\times \sigma$ emission). The gray scale shows HST/WFC3 F160W cutouts of the field from the CANDELS survey \citep{grogin11,koekemoer11}, centered at the coordinates of the line candidate (each region is $14''\times14''$ wide). {\em Right:} Spectra at the position of the line detections are corrected for primary beam attenuation. Solid histograms with error bar show the observed spectra. Negative line candidates are plotted in violet. The dotted black line shows the zero level, the dashed red line shows the best Gaussian fit (numbers are recorded in Tab.~\ref{tab_lines}). The horizontal green bar shows the channels used in the line fits. }
\label{fig_cproc_spc}
\end{figure*}

\subsection{Follow--up observations}\label{sec_followup}

In addition to the 3\,mm scan, we obtained additional observations at the PdBI in the 2mm band to cover a higher--$J$ CO line of the brightest line candidate in our 3\,mm scan (ID.03, below). These observations were obtained between December 15, and December 28, 2011 (4 tracks), with the array in 6-antenna compact configuration (`special'). Baselines ranged between 19.0\,m and 98.1\,m. The pointing center was RA=12:36:49.10, Dec=+62:12:11.3 (J2000.0), i.e. $5.6''$ away from the position of the 3\,mm line. The observing frequency was tuned to 166.500 GHz, encompassing the CO(4-3) transition at $z$=1.784. The final cube includes 7469 visibilities, corresponding to 6.22\,hr on source. The beam is $3.5''\times2.6''$, and the noise per 90 \kms{} channel is 0.96 mJy\,beam$^{-1}$.  In addition, the CO(1-0) transition of this line candidate was targeted with the JVLA in Q band (41.39\,GHz) in C--array, resulting in a resolution of $0.6''$ (noise of 0.09\,mJy over a channel width of 45.5\,MHz, i.e. the full CO(2-1) line width). These observations are discussed in \S\ref{sec_notes} (ID.03) and the corresponding figure (Fig.~\ref{fig_al1}).

\section{Line search approach}\label{sec_method}

The main goal of this project is to blindly search for line emission in our scan, with no pre--selection based on optical/NIR observations. In the following we discuss the line--searching algorithms used in this analysis.

\subsection{The {\em spread} analysis}\label{sec_spread}

Our first method is based on the noise statistics in each channel. Typically, pixel values in each channel map have a Gaussian distribution, the width of which is set by the noise level at that frequency\footnote{Formally, the pixels are spatially correlated. E.g., a single, highly-deviating visibility would produce regular stripes in the image, thus resulting in an excess of bright positive and negative pixels. Moreover, our resolution element is $\sim4$ times larger than the pixel size. Nevertheless, the analysis described here does not rely on a probabilistic interpretation of the pixel distribution, and is only marginally affected by the number of pixels per resolution element. }. If a bright, point-like source is present at a given frequency (redshift), a few pixels will show a clear deviation from the Gaussian distribution, thus introducing a wing towards high signal--to--noise ratios (SNR) of the distribution. We quantify this deviation by defining a `spread' parameter:
\begin{equation}\label{eq_spread}
{\rm spread} = \Sigma_0^{\pm\infty} f_{\rm obs}({\rm SNR})/f_{\rm exp}({\rm SNR}).
\end{equation}
Here $f_{\rm obs}$ is the observed distribution of pixel values, in units of SNR, while $f_{\rm exp}$ is the theoretical gaussian (i.e., a gaussian with $\sigma$=1 and an integrated area equal to the number of pixels). In this step we consider only pixels within $\sqrt{2}\,\times$ the primary beam's half width at half maximum (i.e., twice the area of the primary beam). The sum is computed from zero to $+\infty$ (i.e., tracing positive excesses) and from zero to $-\infty$ (i.e., tracing negative excesses). The latter is used to evaluate the purity of our selection (while positive excesses may be attributed either to real sources or noise, negative excesses are only due to noise). Due to its formulation, the {\em spread} index defined in equation \ref{eq_spread} is formally reminiscent of Pearson's $\chi^2$. We regularize channel-by-channel variations by applying a box-car smoothing over 3 channels ($270$ \kms{}) before computing the {\em spread} index. Figure~\ref{fig_spread} shows an example of a cube with a clear line detection (left panel), and of an `empty' cube with no significant line candidate (right panel). 

\begin{figure*}
\begin{center}
\includegraphics[width=0.80\textwidth]{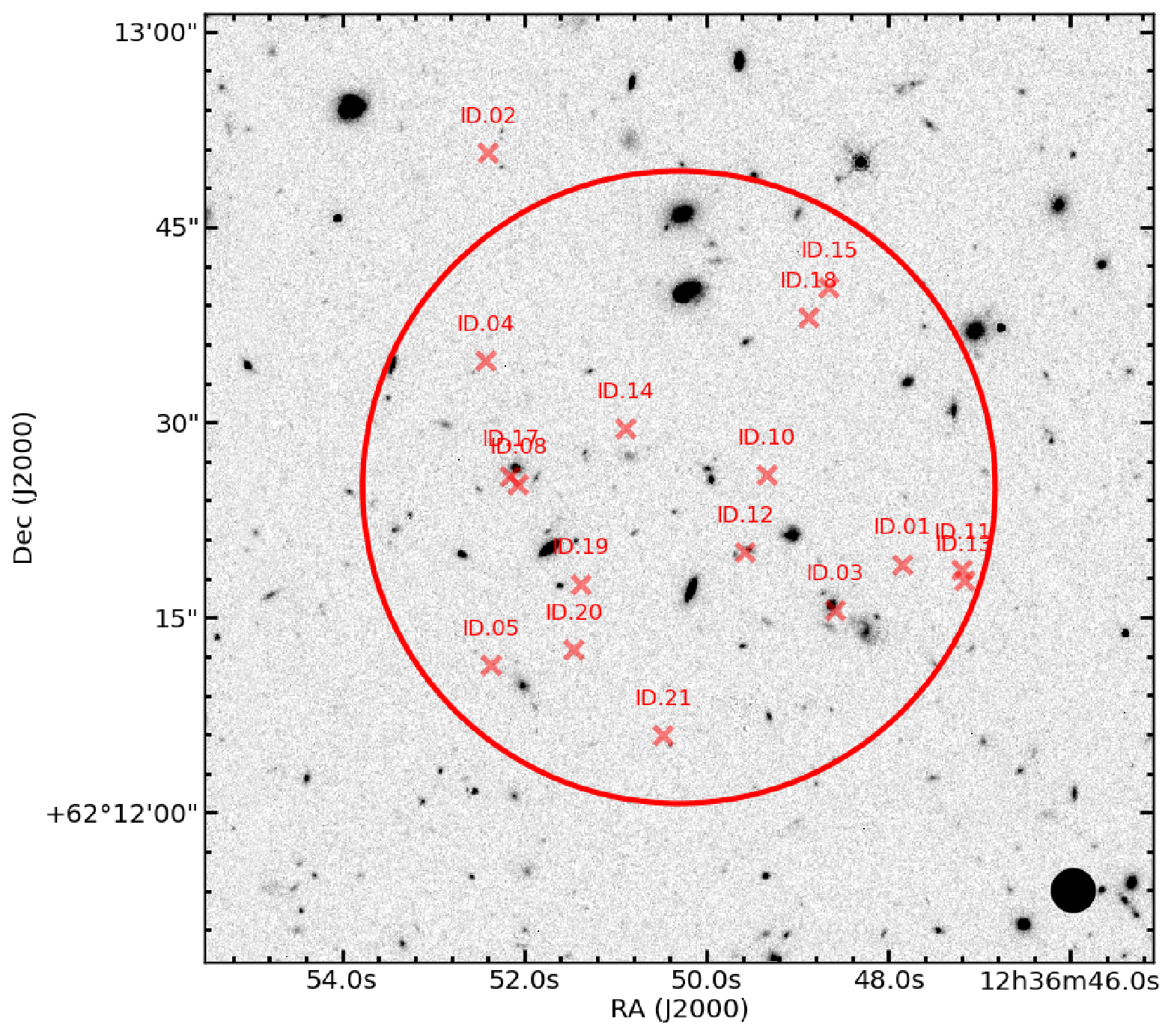}\\
\end{center}
\caption{HST/WFC3 F160W (1.6 $\mu$m, greyscale) image of the region of the HDF--N studied here, from the CANDELS survey \citep{grogin11,koekemoer11}. The big circle shows the primary beam of our observations at the central frequency of our scan (97.25 GHz). The positive line candidates discovered in our searches are plotted with red crosses (from Tab.~\ref{tab_lines}). The typical beam size of our observations is shown as a filled black circle in the bottom right. }
\label{fig_field}
\end{figure*}
Our {\em spread} analysis is a simple but effective hypothesis test. Compared with other statistical tools (e.g., $\chi^2$ test, Kolmogorov-Smirnov, Cram\'{e}r-von Mises, etc), the {\em spread} index has the advantange of down-weighting the central part of the distribution (i.e., many pixels with a measured flux around zero) and highlighting the presence of outliers in the pixel value distribution. Its applicability is limited to data sets for which we have an accurate analytical description of the noise properties ($f_{\rm exp}$). Such a requirement is generally satisfied in our data set (only a couple of channels over the whole scan are affected by receiver `parasites', which introduce strong non-gaussian components in the noise pattern). However, this method is not sensitive to the position of the excess pixels in the map as no spatial coherence is required. For instance, a peak in the {\em spread} distribution as a function of the observed frequency may be associated with a single source (as in the case shown in Figure \ref{fig_spread}, left), or two sources, or more. We have visually inspected all the peaks revealed by the {\em spread} analysis, and progressively lowered the threshold above which a peak in the {\em spread} distribution was considered ({\em spread}=15). The same threshold has been adopted for negative candidates\footnote{We note that, since we are dealing with dirty cubes, bright sources would introduce both positive and negative pixels. However, even in cases of modest (U,V) coverage, negative pixels due to secondary lobes in the beam shape are typically much less prominent (by a factor $\gsim3$) than the central positive peak associated with a point source. Therefore, their imprint on the {\em spread} distribution will be minimal and will not affect our search.}.

\subsection{The \textsf{cprops} search}

In parallel, we have run a different line--searching algorithm based on the finder \textsf{cprops} which has originally been designed to identify giant molecular clouds in nearby galaxies \citep{rosolowski06}. This algorithm first computes the noise in the maps through the data cube, and divides the observed fluxes by the same value, so that data are stored in units of SNR. A mask is associated with any pixel exceeding a given SNR threshold in two consecutive channels. This mask looks for the `core' in the profiles of line candidates. A second mask is generated by selecting the pixels that have a positive excess (down to a lower threshold) in two neighboring channels. This second mask thus identifies the wings of the lines. If the masks overlap in more than two consecutive channels, the excess is considered significant. Compared with the {\em spread} analysis, this approach has the advantage of using the spatial information available, but its purity is usually lower, unless very conservative significance cuts are adopted. Another difference is that the \textsf{cprops} search implicitly assumes Gaussian properties of the noise, while the {\em spread} analysis is in principle applicable to any distribution of data for which an analytical model of the noise is available.

\subsection{Selection of line candidates}\label{sec_candidates}

We ran the \textsf{cprops} algorithm by requiring a $>3$-$\sigma$ excess for the line core, and a $>1.5$-$\sigma$ excess in the neighboring channels. This search produced 117 positive line candidates and 101 negative ones. We visually inspected all of them, and cross-matched them with the results from the {\em spread} analysis. A large number of the sources were excluded because 1) the peak position observed in one channel drifts by a few arcsec in the neighboring channels; or 2) the {\em spread} analysis does not reveal any significant excess. The final list of candidates includes only sources which have been selected by both methods. Our compilation thus consists of 17 positive and 4 negative line candidates (see Tab.~\ref{tab_lines} and Fig.~\ref{fig_cproc_spc}). Since our analysis deals with positive and negative line candidates in the same way, we expect about 4 false positive candidates out of 17. The location of the line candidates in the HDF--N is shown in Fig.~\ref{fig_field}.

Lentati et al.~(in prep.) have developed a completely independent line searching technique which relies on Bayesian inference. This method computes the probability that a given feature in the data is a real galaxy, based on a source model (e.g., expected line profiles, physical sizes, fluxes). 
This search confirmed the detection of all the brightest ($>0.4$ \jykms{}) lines identified in our search. The most significant line candidate found by Lenatati et al.\ that is not included in the present paper is at RA=12:36:48.32, Dec=+62:12:38.44, $\nu=89.28$ GHz. This line candidate is observed in the {\em spread} analysis, but is not selected by \textsf{cprops} since in no individual channel the line candidate exceeds the $3$-$\sigma$ threshold. For consistency we exclude this candidate in the remainder of our analysis.
	 
\begin{table*}
\caption{{\rm Catalogue of the line candidates identified in our analysis, ordered by frequency. (1) Line ID. (2-3) Right ascension and declination (J2000).  (4) Peak frequency and uncertainty, based on Gaussian fit. (5) Integrated flux and uncertainty. (6) Signal-to-noise ratio of the line, as measured on the collapsed map of the line. (7) Signal-to-noise ratio of the line, as measured from gaussian fits. (8) Line Full Width at Half Maximum, as derived from gaussian fit. (9) Quality flag of the line candidates: 1- high quality (SNR$_{\rm spec}>3.5$); 2- low quality (SNR$_{\rm spec}<3.5$). (10) Notes.}} \label{tab_lines}
\begin{center}
\begin{tabular}{cccccccccc}
\hline
ID  & RA          & Dec         & Frequency       & Flux             & SNR$_{\rm map}$ & SNR$_{\rm spec}$ & FWHM & Quality & Notes  \\
    & (J2000.0)   & (J2000.0)   & [GHz]           & [Jy km s$^{-1}$] &                 &                  &[\kms]& 	   &        \\
 (1) & (2)        & (3)         & (4)             & (5)              & (6)             & (7)              & (8)  &  (9)    & (10)   \\
\hline
ID.01     & 12:36:47.72 & +62:12:19.0 & $ 80.05\pm0.03$ & $ 0.531\pm0.116$ & 5.0 &  4.6 & $440_{-200}^{+250}$ & 1 & High--quality \\  
ID.02     & 12:36:52.27 & +62:12:51.7 & $ 82.07\pm0.02$ & $ 0.374\pm0.106$ & 4.3 &  3.5 & $130_{-40}^{+140}$  & 1 & High--quality \\  
ID.03     & 12:36:48.58 & +62:12:15.5 & $ 82.80\pm0.02$ & $ 0.502\pm0.080$ & 7.3 &  6.3 & $330_{-40}^{+120}$  & 1 & Secure   \\  
ID.04     & 12:36:52.56 & +62:12:35.0 & $ 84.94\pm0.03$ & $ 0.189\pm0.070$ & 5.1 &  2.7 & $210_{-50}^{+230}$  & 2 &	     \\  
ID.05     & 12:36:52.37 & +62:12:11.1 & $ 89.89\pm0.04$ & $ 0.473\pm0.112$ & 4.6 &  4.2 & $360_{-40}^{+80}$   & 1 & High--quality \\  
ID.06     & 12:36:49.91 & +62:12:27.2 & $ 90.01\pm0.03$ & $-0.227\pm0.067$ & 4.6 &  3.4 & $240_{-140}^{+150}$ & 2 & Negative \\  
ID.07     & 12:36:47.04 & +62:12:41.7 & $ 91.84\pm0.04$ & $-0.378\pm0.146$ & 3.7 &  2.6 & $290_{-40}^{+40}$   & 2 & Negative \\  
ID.08     & 12:36:52.15 & +62:12:25.8 & $ 93.17\pm0.03$ & $ 0.423\pm0.097$ & 4.4 &  4.4 & $490_{-50}^{+200}$  & 1 & HDF850.1 \\  
ID.09     & 12:36:51.15 & +62:12:40.6 & $ 93.91\pm0.04$ & $-0.194\pm0.073$ & 4.5 &  2.7 & $120_{-50}^{+150}$  & 2 & Negative \\  
ID.10     & 12:36:49.22 & +62:12:25.7 & $103.88\pm0.03$ & $ 0.188\pm0.044$ & 4.0 &  4.2 & $150_{-50}^{+50}$   & 1 & High--quality \\  
ID.11     & 12:36:47.10 & +62:12:18.0 & $108.32\pm0.03$ & $ 0.362\pm0.101$ & 4.0 &  3.6 & $150_{-40}^{+70}$   & 1 & High--quality \\  
ID.12     & 12:36:49.57 & +62:12:19.9 & $108.43\pm0.04$ & $ 0.262\pm0.074$ & 4.2 &  3.5 & $210_{-50}^{+150}$  & 1 & High--quality \\  
ID.13     & 12:36:47.19 & +62:12:18.6 & $108.77\pm0.03$ & $ 0.242\pm0.115$ & 4.1 &  2.1 & $140_{-50}^{+50}$   & 2 &	     \\  
ID.14     & 12:36:50.85 & +62:12:30.1 & $109.65\pm0.03$ & $ 0.244\pm0.072$ & 5.3 &  3.4 & $190_{-40}^{+50}$   & 2 &	     \\  
ID.15     & 12:36:48.56 & +62:12:40.4 & $109.79\pm0.03$ & $ 0.459\pm0.113$ & 4.6 &  4.1 & $260_{-50}^{+120}$  & 1 & High--quality \\  
ID.16     & 12:36:50.49 & +62:12:16.1 & $110.55\pm0.04$ & $-0.311\pm0.087$ & 3.6 &  3.6 & $240_{-40}^{+150}$  & 1 & Negative \\  
ID.17     & 12:36:51.99 & +62:12:25.6 & $111.85\pm0.06$ & $ 0.366\pm0.101$ & 5.4 &  3.6 & $370_{-120}^{+150}$ & 1 & HDF850.1 \\  
ID.18     & 12:36:48.79 & +62:12:38.0 & $112.63\pm0.04$ & $ 0.584\pm0.115$ & 5.6 &  5.1 & $290_{-40}^{+80}$   & 1 & High--quality \\  
ID.19     & 12:36:51.37 & +62:12:16.9 & $113.45\pm0.04$ & $ 0.436\pm0.116$ & 4.3 &  3.8 & $350_{-40}^{+150}$  & 1 & Secure   \\  
ID.20     & 12:36:51.54 & +62:12:12.0 & $113.45\pm0.05$ & $ 0.388\pm0.109$ & 3.6 &  3.6 & $150_{-70}^{+100}$  & 1 & High--quality \\  
ID.21     & 12:36:50.48 & +62:12:05.8 & $114.15\pm0.05$ & $ 0.360\pm0.132$ & 3.2 &  2.7 & $220_{-90}^{+180}$  & 2 &	     \\  
\hline
\end{tabular}
\end{center}
\end{table*}								 

\subsection{Completeness and purity from artificial data cubes}\label{sec_complete}

\begin{figure}
\begin{center}
\includegraphics[width=0.99\columnwidth]{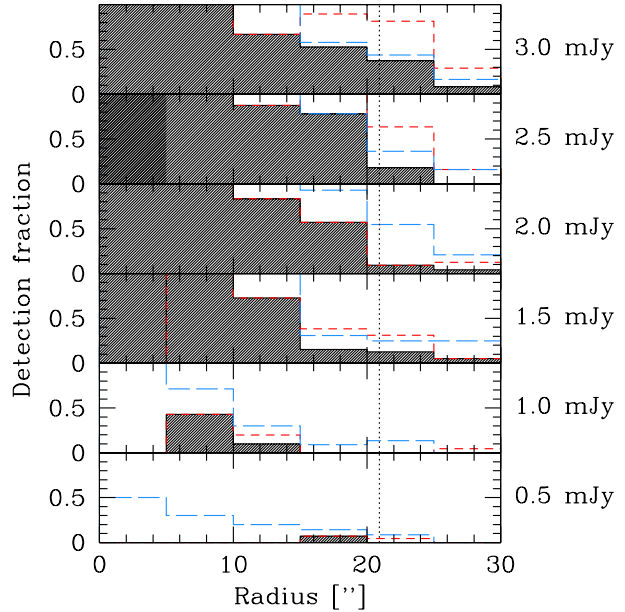}\\
\end{center}
\caption{Completeness (detection fraction) of our line searching techniques as a function of input source flux and offset with respect to the pointing center. The shaded histograms show the fraction of input sources that we successfully recovered with the combined use of the {\em spread} analysis and with \textsf{cprops}. The completeness levels achieved by applying only one of the two approaches are shown as empty histogram (blue, long-dashed = {\em spread}, red, short-dashed = \textsf{cprops}). The vertical dotted line shows the primary beam radius at the central frequency of the test cube. At the level of flux densities of interest here, we reach $>66$\% completeness.}
\label{fig_completeness}
\end{figure}

In order to estimate the completeness, degree of purity, and role of the Eddington bias in our line searches, we generated data cubes from artificial visibilities with the same format (i.e., UV-coverage, spectral setup, number of visibilities, etc) of a typical data cube in our observations. We included Gaussian noise on the real and the imaginary parts of each visibility, resulting in a rms of 0.3 mJy\,beam$^{-1}$ per 90\,\kms{} channel. We then added 10 sources that are randomly distributed over a region twice as big as the primary beam. Each source represents a point source with an emission line peaked at a random frequency within the cube, and no continuum emission. The line profile is assumed to be Gaussian, with a FWHM=300\,\kms{}, and peak flux set to 3.0 mJy (scaled down according to the primary beam attenuation). We generated 10 realizations of each cube (i.e., 100 sources in total). We then ran our line searching algorithms, and counted the number of input sources that have been successfully recovered. The whole process was repeated for sources with peak flux = 2.5, 2.0, 1.5, 1.0, and 0.5\,mJy. Fig.~\ref{fig_completeness} shows the results of this analysis. We find that $>$50\% of the input sources within the primary beam are recovered for line peak fluxes of 1.5\,mJy. The completeness increases at decreasing distance from the pointing center, and at increasing fluxes: $\sim66$\% at 1.5--2.0\,mJy (comparable with the typical fluxes of the sources in Table~\ref{tab_lines}) and reaches up to $85$\% for line peak fluxes $\gsim2.5$ mJy. 

In each realization, we then counted the number of sources identified by the {\em spread} analysis, by \textsf{cprops}, and by both, which do not match any input source. These detections allow us to estimate the rate of false positives in our sample. We find that, on average, the {\em spread} analysis identifies 5.2 false positives per cube, \textsf{cprops} about 2.5, and the combination of the two only 0.1 false positive per cube. This implies that, out of the whole scan (equivalent to 20 cubes), we expect $\sim2$ false positives. This estimate is in agreement with the number of negative peaks identified as potential sources in \S\ref{sec_candidates}.

\section{Results}\label{sec_results}

\subsection{Properties of 3\,mm-selected line candidates}\label{sec_lines}

Fig.~\ref{fig_cproc_spc} shows the line spectra and maps of the 21 line candidates discovered in our study (including `negative' candidates, which are used to gauge the purity of our line search). Line maps are extracted by integrating over the channels encompassing the full width at zero flux of each line candidate. We compute the map--based SNR of each line as the ratio between the line flux (i.e., the value of the brightest pixel in the line map, in Jy\,beam$^{-1}$) and the rms as measured in the line map (see Tab.~\ref{tab_lines}). Then, for each line, we extract the spectra at the coordinates corresponding to the brightest pixel of each line candidate\footnote{We note that this approach may slightly underestimate the significance of real spectral features, because of the (relatively big) uncertainties in the coordinates of line baricenters.}. Spectra are then corrected for primary beam attenuation, assuming a gaussian attenuation and the frequency dependence of the primary beam as described in \S\ref{sec_observations}. The SNR of the line candidates is usually modest, therefore a proper characterization of the line profiles is not possible. In most of the cases, however, a Gaussian fit provides a good description of the observed lines. We assume no underlying continuum emission, justified by the analysis of our continuum image (see \S\ref{sec_cont}). The measured line fluxes range between 0.16 and 0.56 \jykms{}. Line widths, as derived from the Gaussian fits, range between 110 and 500 \kms{} with significant uncertainties. The spectrum--based SNR is computed as the ratio between the fitted line flux and its uncertainty\footnote{SNR$_{\rm spec}$ is lower than the map--based SNR, because of model uncertainties (lines not being well described by a single gaussian) and because the Gaussian fits also include the line wings, which worsen the noise, without contributing significant flux, while the line map is extracted only in channels where the line is clearly detected.}.

Based on the spectral SNR of the line candidates, we define a quality flag: We consider high--quality candidates (quality flag: 1) objects with SNR$_{\rm spec}>3.5$ (14 candidates), and we label as low-quality candidates (quality flag: 2) those with SNR$<$3.5 (7 candidates). We note that 3 out of 4 of the negative peaks belong to the ``low--quality'' class, while at least 2 lines in the ``high--quality'' class are known to be real (the two CO transitions associated with HDF850.1). 

\begin{figure}
\includegraphics[width=\columnwidth]{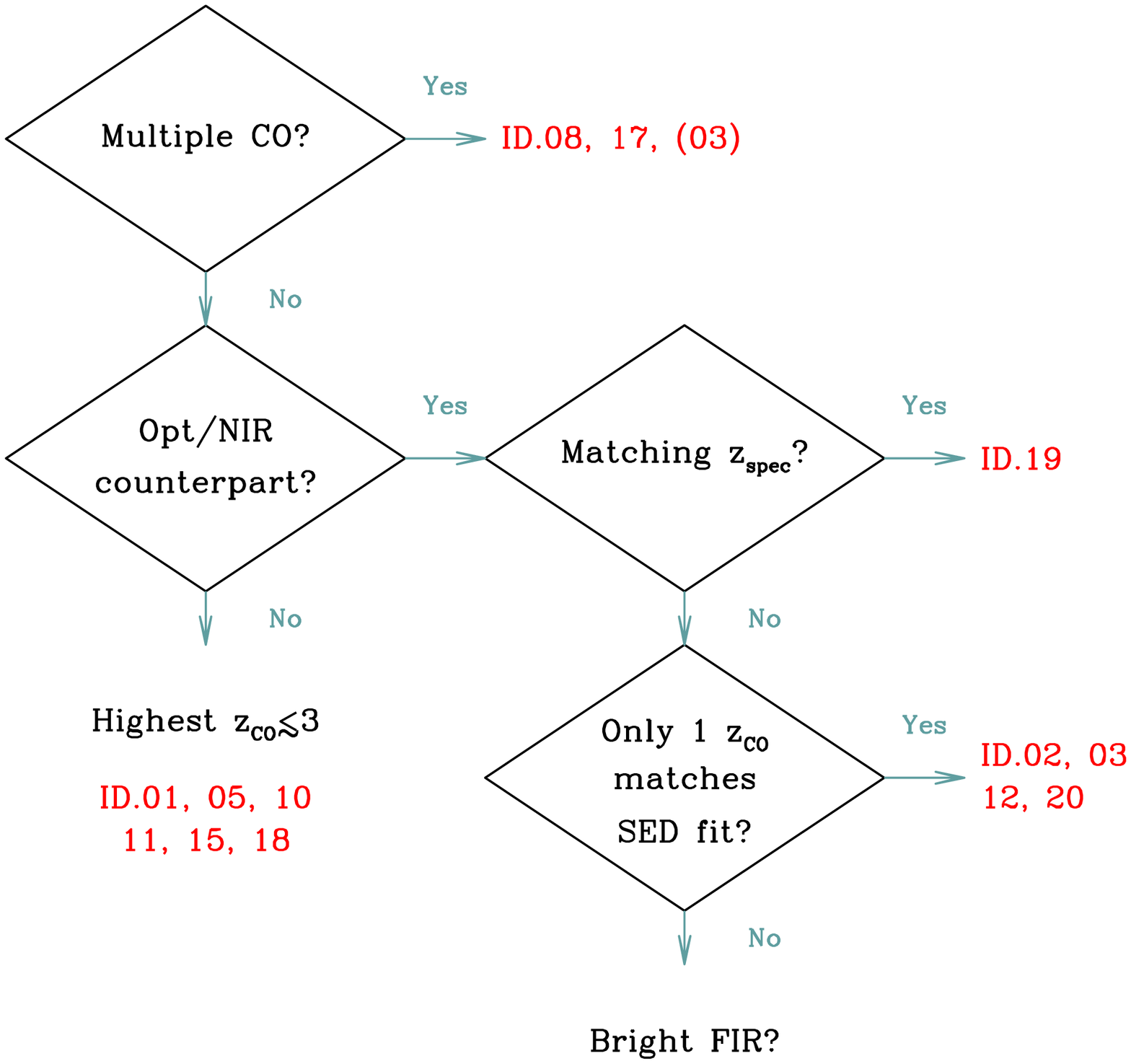}\\
\caption{Scheme adopted in order to assign a CO transition (and thus redshift) to each of the line candidates of the molecular scan. We first look for multiple CO transitions in the 3\,mm scan (implying a redshift $z\gsim3$, see Fig.~\ref{fig_z_range}); then we search for counterparts at optical/NIR wavelengths. If none is found (implying a faint / obscured / high-$z$ candidate galaxy), we assign the highest $z$ consistent with only 1 CO transition in the 3\,mm window to the galaxy. If a counterpart is available, we check whether it has a spectroscopic redshift consistent with $z_{\rm CO}$; if not, we base our CO identification on the goodness of the SED fits. If none of the above applies, we opt for the $z_{\rm CO}$ for which the SED fit predicts the brightest FIR emission. The high--quality line candidates in our study are labeled in red. ID.03 is reported in bracket as a `multiple CO' identification, since the redshift is confirmed by other CO observations (CO(1-0) at 41.4 GHz and CO(4-3) at 165.5 GHz), but they are not part of the original 3\,mm scan (\S~\ref{sec_followup}).}
\label{fig_ident}
\end{figure}

\subsection{Assigning redshifts to line candidates}

The most prominent lines that we expect to detect in the 3\,mm band are various CO transitions (up to $J_{\rm up}\sim7$) and the neutral carbon fine-structure lines \Ci{}$_{1-0}$ and \Ci{}$_{2-1}$ (see Tab.~\ref{tab_z_range} and Fig.~\ref{fig_z_range}). In the following subsections, we assign the most likely redshift (and thus CO transition) for each line candidate. In Fig.~\ref{fig_ident} we provide a sketch of our approach to identify which CO transition (and thus redshift) the respective line candidates correspond to. First, we look for multiple transitions arising at the same spatial position (which would immediately constrain the redshift). We then search for an optical/NIR counterpart in the deep ancillary database of the HDF--N. If none is found, we identify the line candidate as the highest--$J$ CO transition consistent with the lack of multiple lines in the 3\,mm window (i.e., we locate the line at the highest $z_{\rm CO}<3.016$). If one or more optical/NIR counterparts are available, we check whether a spectroscopic redshift is available and consistent with the line frequency observed in the 3\,mm scan. If no spectroscopic information is available, we select plausible CO identifications on the basis of the UV-to-FIR spectral energy distributions (SEDs) of the counterparts. In the case of remaining ambiguity, we choose to adopt the CO redshift for which the SED fitting predicts the brightest FIR emission. This is justified by the observed correlation between FIR and CO luminoisity \citep[see, e.g.,][]{carilli13}: The SED model predicting the brightest dust emission also predicts brighter CO emission, which would increase the chances to detect a line in our scan. In the following sections, we discuss in more details the individual steps of this approach. 

\subsection{Search for multiple CO transitions}
\begin{figure*}
\includegraphics[width=0.49\textwidth]{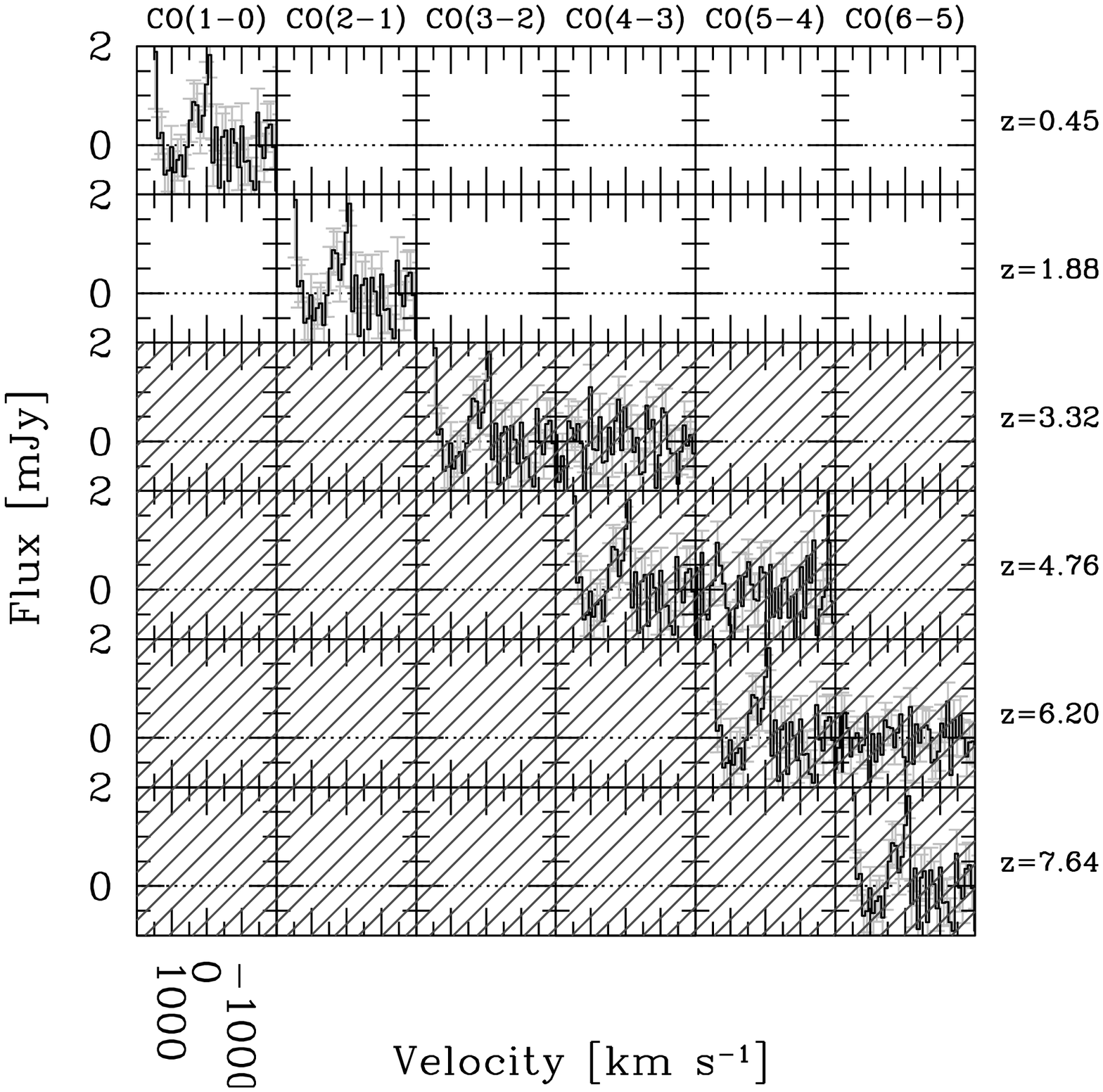}
\includegraphics[width=0.49\textwidth]{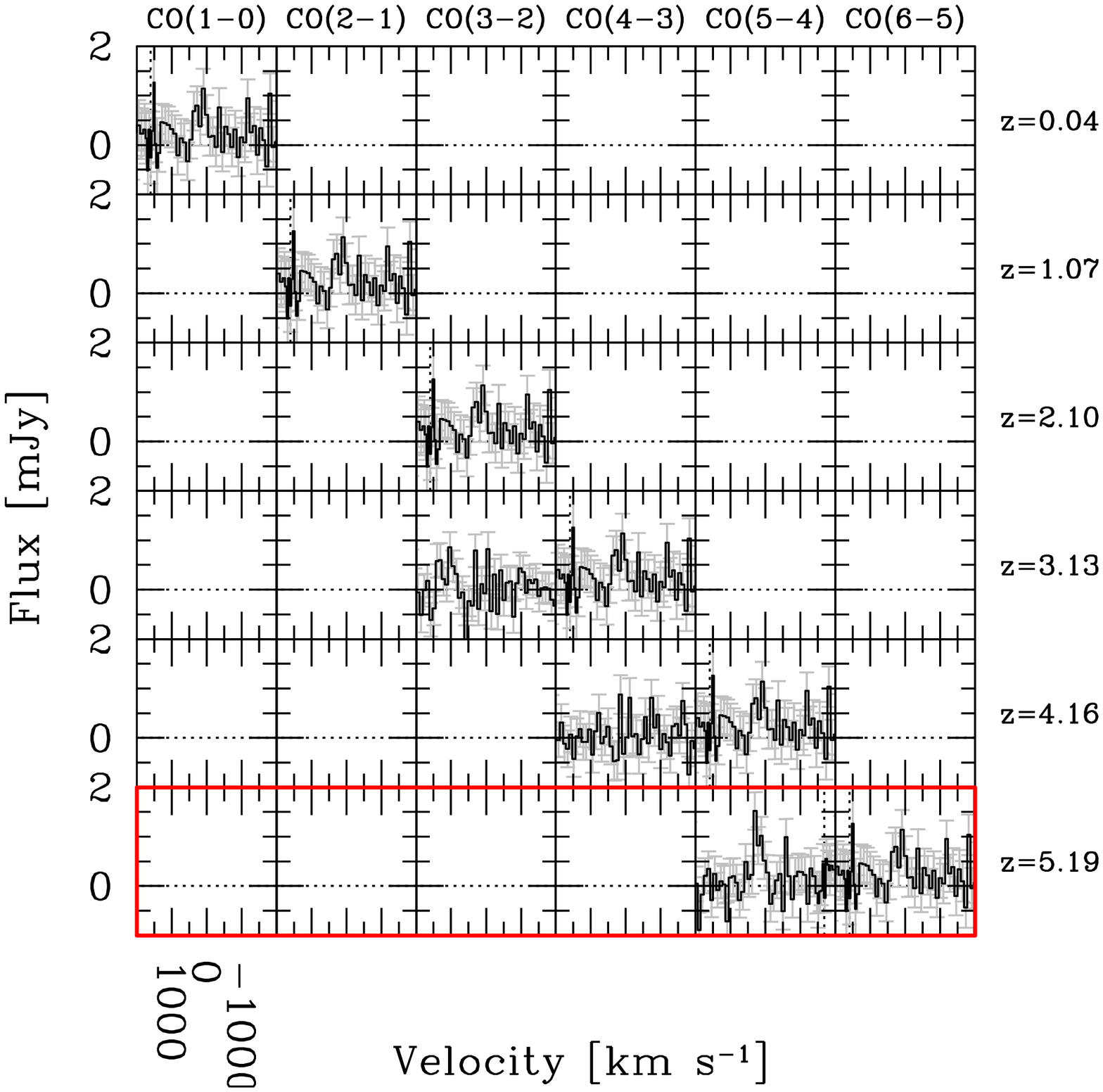}\\
\caption{Diagnostic for line identification for two of the most prominent lines discovered in our scan. If the line is identified with a given CO or \Ci{} transition, we examine which other lines do we expect to see, and at which redshift would the source be. In the left panel, ID.01 shows no other obvious line, suggesting that this source is at $z<3$. On the other hand, ID.17 (right panel) shows two lines (the other one being ID.08 in our analysis) consistent with CO(5-4) and CO(6-5) at $z\approx5.19$ \citep{walter12}.}
\label{fig_which_co}
\end{figure*}

If a source is at $z>3.016$, at least two CO or \Ci{} lines are covered, providing an accurate redshift\footnote{We note that some ambiguity may persist at these low SNRs, as CO transitions have frequencies which are $\nu_0$[CO($J_{\rm up}$-$J_{\rm up-1}$)]$\approx J_{\rm up} \times \nu_0$[CO(1-0)]. For instance, two lines identified as CO(6-5) and CO(4-3) at $z_1$ can also be identified as CO(3-2) and CO(2-1) at $z_2=(z_1-1)/2$. However, in the $z_2$ scenario the CO(5-4) line should also be detected, thus allowing us to discriminate between the two cases.}. In Fig.~\ref{fig_which_co} we show our basic diagnostic diagram for line identification, applied to two line candidates. In one case (line candidate ID.01), there is only one line observed in the 3\,mm scan, thus excluding any $z>3$ identification (unless an extremely steep CO excitation ladder is invoked). The second case shown in Fig.~\ref{fig_which_co} has two lines (both independently identified in the line search process described in \S\ref{sec_method}: lines ID.08 and ID.17). These are associated with the SMG HDF\,850.1 and they are consistent with being CO(5-4) and CO(6-5) at $z\approx5.18$ \citep{walter12}. HDF\,850.1 is the only source for which we detected more than one line in the molecular line scan. Another pair of line candidates (ID.11 and ID.13) are also spatially consistent, but their frequency difference cannot be explained by any combination of CO or \Ci{} lines at a single redshift. Therefore, we will treat these two lines as independent sources. All the remaining candidates show only one transition in the scan, thus suggesting that they are at $z\lsim3$.

\subsection{Optical/NIR counterparts of CO line candidates}\label{sec_counterparts}

As a next step we search for counterparts at optical/NIR wavelengths of the CO line candidates. We take the HDF--N catalogue of galaxies as reference \citep{williams96}, and we refer to the compilation of photometric redshifts by \citet{soto99}. We also compare the positions of the line candidates in our survey with the IRAC and MIPS maps of this region (PI: M.~Dickinson). Given the modest SNR of our line detections, the uncertainties on the spatial position of the sources are comparable to the size of the resolution element (i.e. a radius of $\sim1.5''$). Here we consider as tentative counterparts those galaxies that are within $1.5''$ of the peak position of the CO line candidates. Ten line candidates have 1 or more associations in the catalogue by \citet{soto99}. One of them is a negative feature in our 3\,mm scan (ID.06) and will be ignored in this analysis. Both ID.08 and ID.17 (the two transitions identified with HDF850.1) are spatially consistent with a $z\approx1$ galaxy; however, the redshift inferred by our 3\,mm scan rules out such an association, as confirmed by the \Cii{} observations at higher spatial resolution presented in \citet{walter12}.

The remaining 7 CO line candidates have reliable galaxy associations. We collect photometric information of these sources from the catalogues by \citet{williams96} ($ubvi$), \citet{dickinson03} ($H$ band), \citet{bundy09} ($bvizK$), Dickinson et al.\,(in prep.) (IRAC 3.6 $\mu$m, 4.5 $\mu$m, 5.8 $\mu$m, 8.0 $\mu$m and MIPS 24 $\mu$m) and \citet{elbaz11} (Herschel/PACS 100 $\mu$m, 160 $\mu$m and SPIRE 250 $\mu$m, 350 $\mu$m, 500 $\mu$m). We also include limits on the 3\,mm continuum emission, as derived from the map presented in \S\ref{sec_cont}. We fit these spectral energy distributions (SEDs) using \textsf{magphys} \citep{dacunha08,dacunha13}. This physically--motivated model combines the attenuated stellar emission from the UV to the NIR with the dust emission in the MIR to sub-mm assuming energy balance between the radiation absorbed at UV and optical wavelengths and the one re-radiated by dust at IR wavelengths. We shift galaxy templates to the redshifts corresponding to possible CO identifications of our line candidates. The available photometric information on the optical/NIR counterparts allow us to exclude some possible CO identifications. The potential counterparts of our CO line candidates have modest stellar masses: $M_*\lsim10^{10}$ \Msun{}. The only exception is the counterpart of ID.03 ($M_*=1.3\times10^{11}$ \Msun{}, see discussion below).

\begin{table*}
\caption{\rm Identification of the line candidates in our analysis, and tentative optical/NIR counterparts. (1) Line ID. (2) Is there any optical/NIR counterpart? (3--4) Coordinates of the optical/NIR counterpart(s). (5) Distance between the peak of the line candidate and its counterpart. (6) Photometric redshift of the optical/NIR counterpart, from \citet{soto99}. (7) Tentative $J_{\rm up}$ identification of the CO transition. Our `second-choice' identification is reported in brackets. (8) Adopted redshift, based on the most likely CO identification (see text for details). (9-10) Line luminosities at the adopted $z_{\rm CO}$. The luminosity numbers are uncertain by a significant factor ($\sim$\,30\%).} \label{tab_zlum}
\begin{center}
\begin{tabular}{cccccccccc}
\hline
ID   & Counter- & RA   & Dec          &Dist.& $z_{\rm phot}$ & $J_{\rm up}$ & $z_{\rm CO}$	 & $L'$	   & $L_{\rm c}$	\\
     & part?        &      &              &[$''$]&     &        &                     & [$10^9$ \Kkmspc] & [$10^6$ \Lsun{}] \\
 (1) & (2)     & (3)  & (4)          & (5) & (6)  & (7)    &     (8)	         & (9)   & (10)   \\
\hline
ID.01 & N &	      & 	     &     &	  & (1),2  & $1.8799\pm0.0012$   & 23.4  &   9.2  \\
ID.02 (1) & Y & 12:36:52.245 & +62:12:51.56 & 1.3 & 0.92 & (1),2  & $1.8084\pm0.0007$   & 15.3  &   6.0  \\
ID.02 (2) & Y & 12:36:52.561 & +62:12:51.56 & 1.5 & 1.12 &	   &			 &	 &	  \\
ID.03 & Y & 12:36:48.626 & +62:12:15.79 & 0.4 & 1.6  &  2     & $1.7844\pm0.0008$   & 20.1  &   7.9  \\
ID.04 (1) & Y & 12:36:52.561 & +62:12:33.61 & 1.5 & 0.64 & (1),2  & $1.7142\pm0.0007$   &  7.0  &   2.8  \\
ID.04 (2) & Y & 12:36:52.485 & +62:12:35.46 & 0.9 & 2    &	   &			 &	 &	  \\
ID.05 & N &	      & 	     &     &	  & (2),3  & $2.8471\pm0.0012$   & 19.0  &  2.52  \\
ID.08 & N &	      & 	     &     &	  & 5	   & $5.1869\pm0.0035$   & 15.9  &  9.71  \\
ID.10 & N &	      & 	     &     &	  & (2),3  & $2.3286\pm0.0008$   &  5.4  &   7.1  \\
ID.11 & N &	      & 	     &     &	  & (2),3  & $2.1924\pm0.0008$   &  9.3  &  1.23  \\
ID.12 & Y & 12:36:49.673 & +62:12:19.69 & 0.8 & 2.08 & 3	   & $2.1891\pm0.0012$   &  6.7  &   8.9  \\
ID.13 & N &	      & 	     &     &	  & (2),3  & $2.1794\pm0.0010$   &  6.2  &   8.2  \\
ID.14 & N &	      & 	     &     &	  & (2),3  & $2.1535\pm0.0009$   &  6.1  &   8.0  \\
ID.15 & N &	      & 	     &     &	  & (2),3  & $2.1497\pm0.0009$   & 11.4  &  15.1  \\
ID.17 & N &	      & 	     &     &	  & 6	   & $5.1856\pm0.0034$   &  9.5  & 101.0  \\
ID.18 & N &	      & 	     &     &	  & (2),3  & $2.0702\pm0.0010$   & 13.6  &  18.0  \\
ID.19 & Y & 12:36:51.602 & +62:12:17.28 & 1.5 & 1.76 & 3	   & $2.0474\pm0.0015$   &  9.9  &  13.1  \\
ID.20 & Y & 12:36:50.396 & +62:12:05.12 & 1.0 & 2.56 & 3	   & $2.0479\pm0.0007$   &  8.8  &  11.7  \\
ID.21 & Y & 12:36:51.358 & +62:12:11.89 & 1.0 & 3.16 & (3),4  & $3.0383\pm0.0021$   &  9.1  &  28.5  \\
\hline
\end{tabular}
\end{center}
\end{table*}								 

\subsection{Notes on individual line candidates}\label{sec_notes}

Hereafter we briefly describe all the positive line candidates in our analysis. We divide the sample in `secure line detections', i.e., line candidates which have been confirmed through corollary observations; `high--quality line candidates', i.e., prominent line candidates which show a spectroscopic SNR$>$3.5 (labeled as quality 1 in Tab.~\ref{tab_lines}); and others. 

\subsubsection{Secure line detections}

\begin{figure}
\includegraphics[width=0.99\columnwidth]{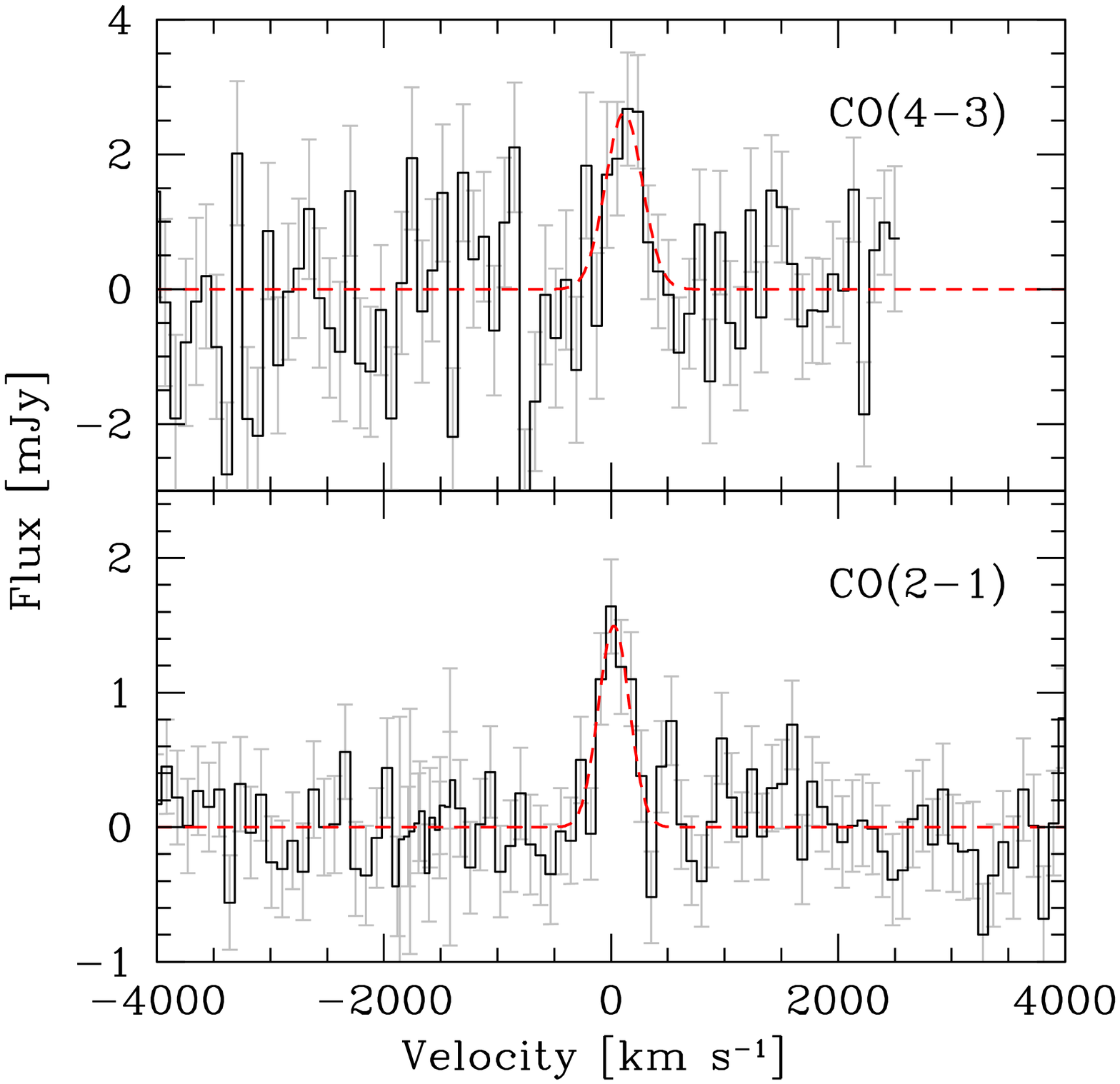}\\
\caption{CO emission of CO(2-1) and CO(4-3) of ID.03 at z=1.784 from our 3\,mm and 2mm PdBI observations. A Gaussian fit gives a peak line flux of $\sim1.5$ mJy and $\sim2.5$ for CO(2-1) and CO(4-3) respectively. The integrated line fluxes are $0.51\pm0.08$ \jykms{} and $1.06\pm0.23$ \jykms{}. The implied luminosities are $L'=2.0\times10^{10}$ \Kkmspc{} and $L'=1.1\times10^{10}$ \Kkmspc{} respectively.}\label{fig_al1}
\end{figure}
\begin{figure}
\includegraphics[width=0.99\columnwidth]{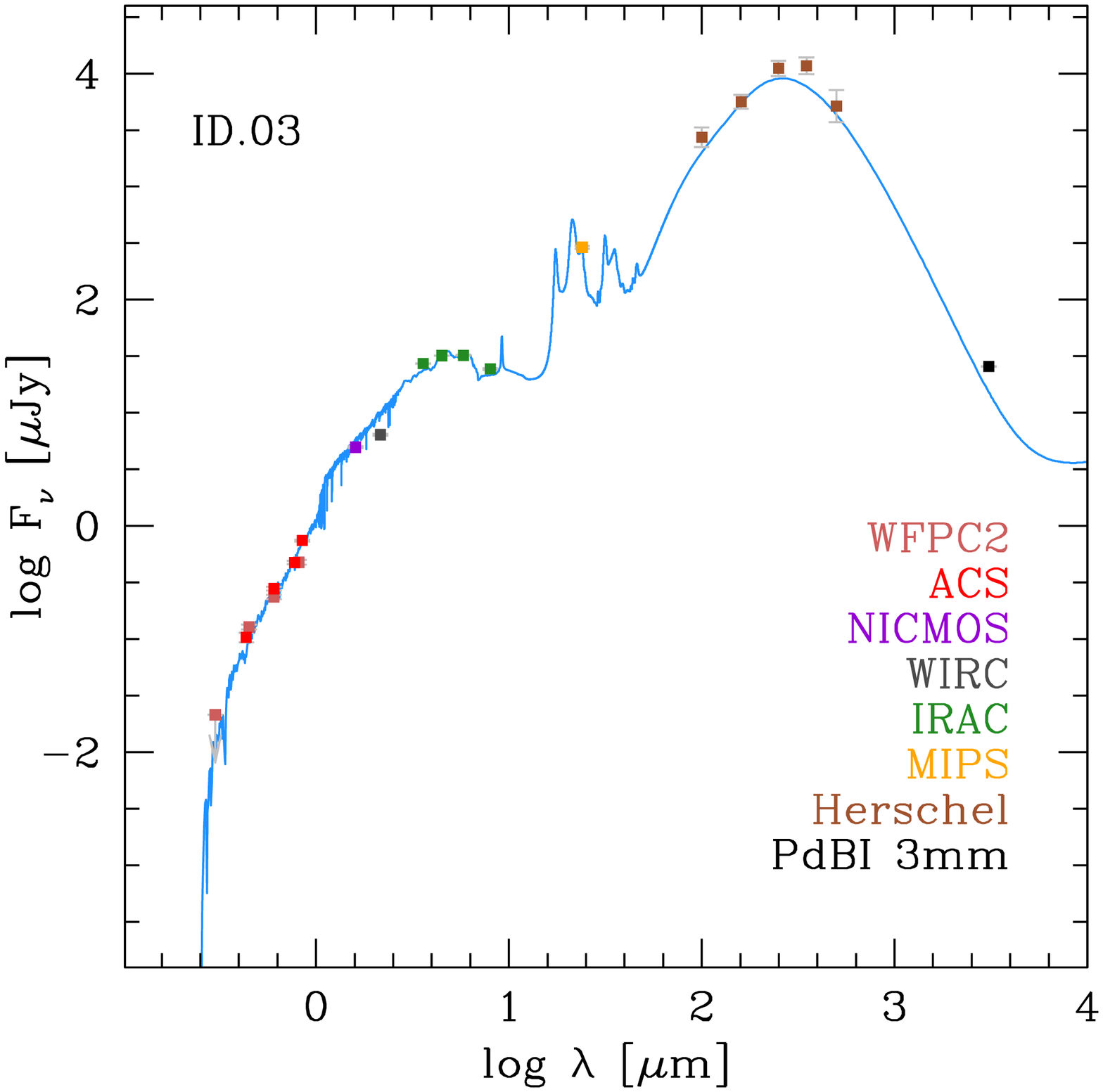}\\
\caption{Spectral energy distribution of ID.03. Observed data are taken from: \citet{williams96} (WFPC2); \citet{bundy09} (ACS+WIRC); \citet{dickinson03} (NICMOS); Dickinson et al.\,in prep. (IRAC+MIPS); \citet{elbaz11} (Herschel) and this work (3\,mm continuum). The fit (shown as a solid line) is performed following \citet{dacunha13}, assuming a redshift of $z=1.784$, as measured based on our CO(2-1) and CO(4-3) detections. In particular, the model of dust emission well fits the observed IR photometry.}\label{fig_sed03}
\end{figure}

\begin{figure}
\includegraphics[width=0.99\columnwidth]{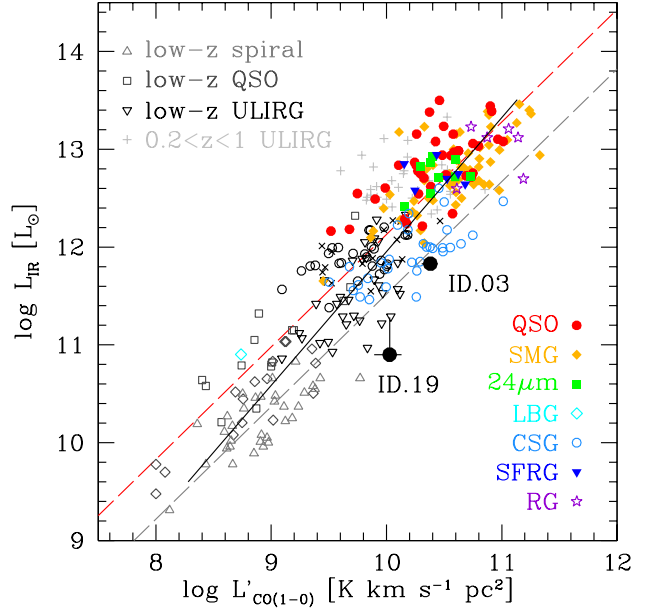}\\
\caption{IR luminosity as a function of CO(1-0) luminosity of the two secure line candidates discovered in our survey, beside HDF850.1. For ID.03, we use the CO(1-0) luminosity extrapolated from the CO(2-1) and assuming a typical SMG CO excitation curve (see text for details). For ID.19, we infer CO(1-0) based on the CO(3-2) luminosity, assuming M82-like excitation. The large uncertainties in IR luminosity for this source are due to the lack of photometric constraints a wavelengths longer than 24 $\mu$m (see Fig.~\ref{fig_sed19}). For a comparison, we plot in rainbow-colored symbols different classes of galaxies at $z>1$, and with gray symbols galaxies at $z<1$, from the compilation by \citet{carilli13}. The two sources discussed in our study both lie below the typical IR--CO(1-0) luminosity relation, but still within the observed scatter, once we account for observational uncertainties.}
\label{fig_co_fir}
\end{figure}

\begin{figure}
\includegraphics[width=0.31\columnwidth]{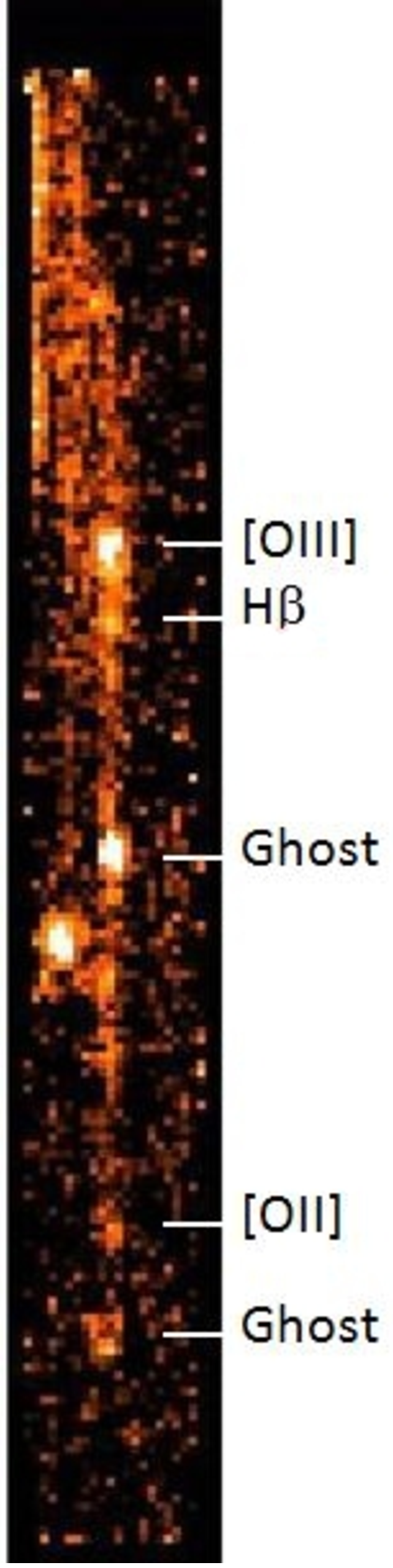}
\includegraphics[width=0.49\columnwidth]{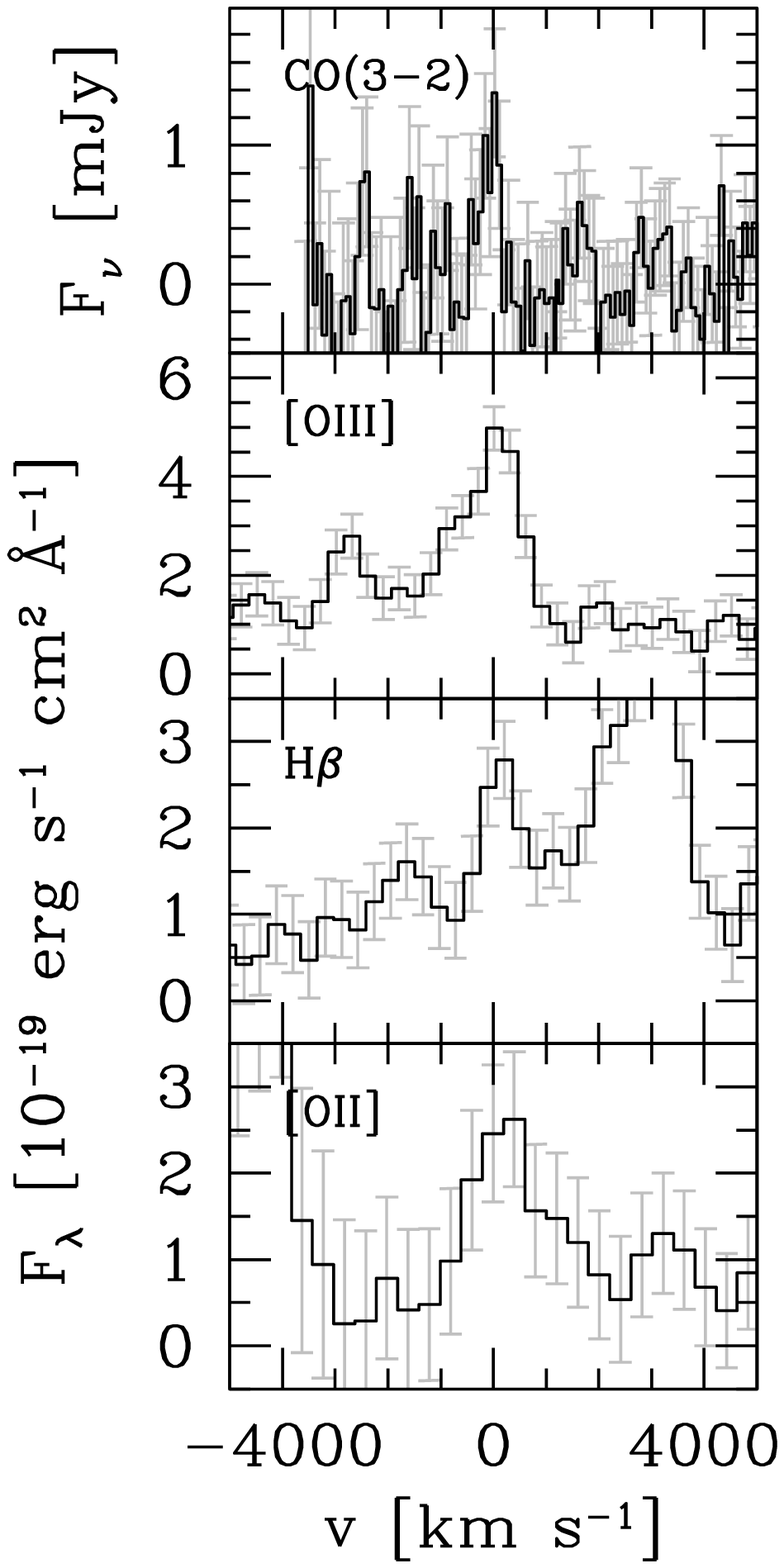}\\
\caption{{\em Left:} Near infrared, 2D grism spectrum of the counterpart associated with the line candidate ID.19, observed with HST as part of the AGHAST survey (Weiner et al.\ in prep.). Wavelength increases from bottom to the top. Main emission lines, identified assuming $z=2.047$, and ghost images from the zero order continuum emission are labeled. {\em Right:} Velocity profile of the CO(3-2) line of ID.19  from our 3\,mm observations (top), and of the rest-frame optical lines shown in the left hand panel. There is excellent agreement in redshift between those lines. We note that rest-frame optical lines appear broadened because of the modest spectral resolution of the grism observations.}\label{fig_id19}
\end{figure}
\begin{figure}
\includegraphics[width=0.99\columnwidth]{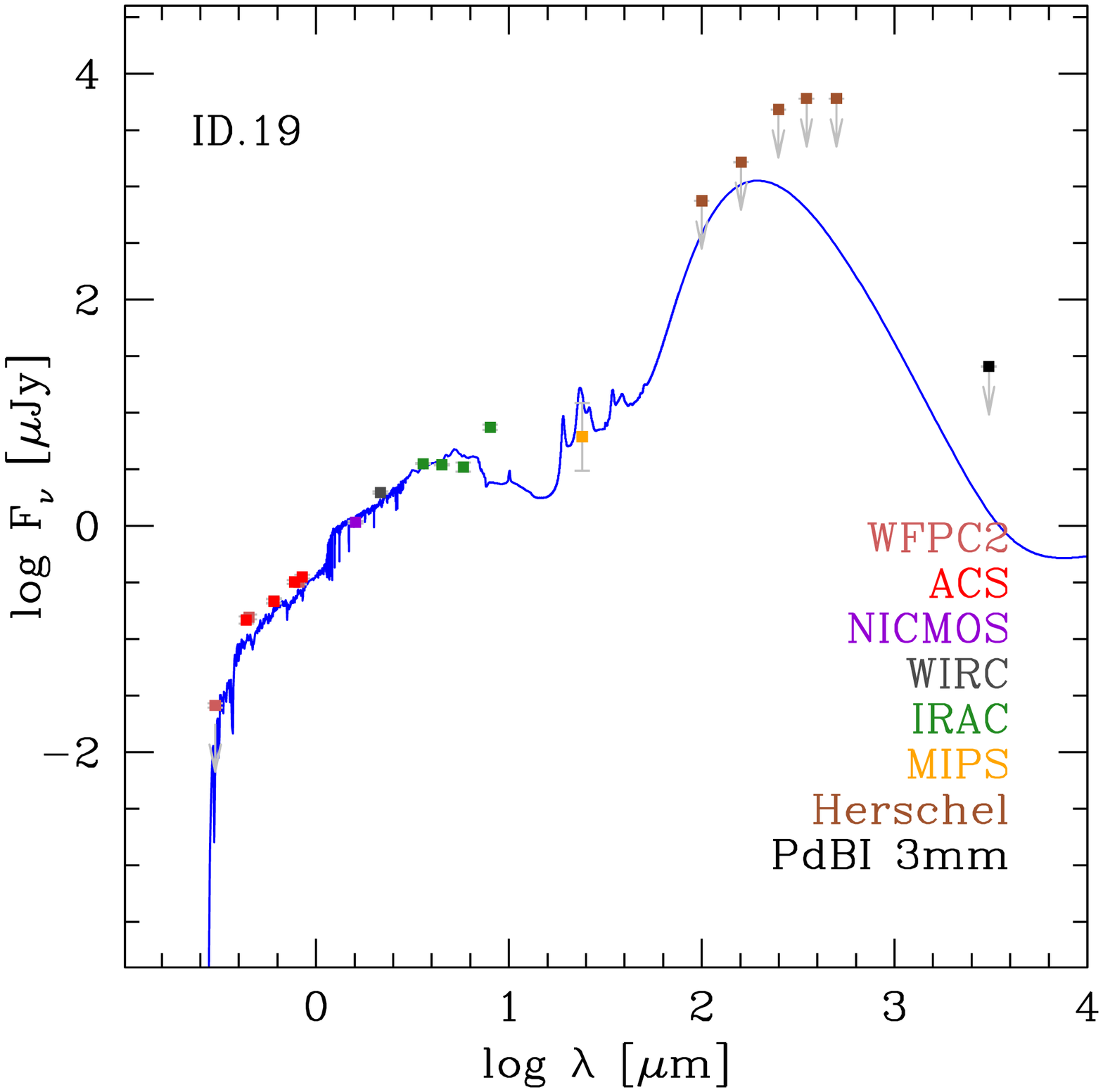}\\
\caption{Spectral energy distribution of ID.19. Symbol code is the same as in Fig.~\ref{fig_sed03}. The fit (shown as a solid line) assumes a redshift of $z=2.047$, based on our CO(3-2) detection (see Fig.~\ref{fig_id19}).}\label{fig_sed19}
\end{figure}
\indent {\bf ID.03} ($\nu=82.80\pm0.02$ GHz) is the brightest line detected in our 3\,mm scan. Our PdBI follow-up 2mm observations reveal a second line at $\approx$165.5\,GHz (Fig.~\ref{fig_al1}). A third, tentative line is reported at $\approx41.4$ GHz in our JVLA observations, albeit only at a SNR of $\sim$3.6$\sigma$. These lines are consistent with being CO(2-1), CO(4-3) and CO(1-0) at $z=1.784$\footnote{Given the relatively low SNR of our detections, the detected lines could in principle also be the CO(4-3), CO(8-7) and CO(2-1) transitions, which would imply a redshift of $z\sim4.5$. However in this case the CO(5-4) line would also be covered by our 3\,mm scan, but is not detected. This leaves $z$\,=\,1.784 as the only plausible redshift.}. The line fluxes are $\sim0.11$ \jykms{}, $0.50\pm0.08$ \jykms{}, and $1.06\pm0.23$ \jykms{} for CO(1-0), CO(2-1), and CO(4-3), respectively. The CO(4-3)/CO(2-1) flux ratio is $\approx2$, i.e., close to the average CO ladder observed in SMGs \citep{carilli13} and significantly higher than what is inferred from the (few) observations of multiple CO transitions in BzK galaxies \citep{dannerbauer09,aravena10}. The observed CO(1-0) flux is also consistent with an SMG-like CO ladder, within the uncertainties. We therefore adopt a line luminosity $L'_{\rm CO(1-0)}=2.4\times10^{10}$ \Kkmspc{}, based not on our quite uncertain measurement of the CO(1-0) line, but on the CO(2-1) luminosity, assuming an $L'_{\rm CO(2-1)}$/$L'_{\rm CO(1-0)}$ ratio of 0.85 \citep{carilli13}. 

ID.03 is coincident with an optically faint but infrared bright galaxy. Its AB magnitudes are: $B=26.4$, $z=24.2$, $K=21.9$ and thus $BzK=(z-K)-(B-z)=0.2$ \citep[i.e. $BzK>-0.2$, the definition of a BzK galaxy following][]{daddi04}. With $(B-z)_{\rm AB}$=2.1 and $(z-K)_{\rm AB}$ = 2.3 this galaxy is also placed inside the box of `star forming galaxies at $z>1.4$' \citep[Fig.~3 in][]{daddi04}. The galaxy is also detected by Herschel, sampling most of the dust continuum emission \citep{elbaz11}. This galaxy has no unambigous spectroscopic redshift, although it has frequently been observed with Keck optical spectrographs\footnote{NIR continuum emission is detected in HST/WFC3 grism observations covering $1.1$--$1.65$ $\mu$m from A Grism H-alpha SpecTroscopy survey (AGHAST; Weiner et al.\ in prep.). No strong emission line, in particular the \Oiii{} emission, is seen, implying that the line is either intrinsically weak or absorbed by dust.}. The WFPC2+NICMOS+K photometric redshift for this galaxy is $z_{\rm phot}\sim1.6$, suggestively close to the CO redshift $z = 1.784$. As shown in Fig.~\ref{fig_sed03}, our CO redshift is in excellent agreement with the observed SED, ranging from the optical HST to Herschel photometry. The source has only a faint VLA 1.4\,GHz detection but is not detected by Chandra and is thus not known to host an AGN. 

Based on the SED fitting we derive the following parameters of the galaxy (median-likelihood estimates; 1--$\sigma$ confidence ranges in brackets): SFR=$38_{-1}^{+8}$ \Msun{}\,yr$^{-1}$; $M_*=(2.5_{-0.3}^{+0.1}) \times 10^{11}$\,\Msun{}; $L_{\rm dust} = (6.8_{-0.16}^{+0.65}) \times 10^{11}$\,L$_\odot$; M$_{\rm dust} = (4.6_{-3.0}^{+10}) \times 10^{7}$\,\Msun{}. Fig.~\ref{fig_co_fir} shows the location of ID.03 with respect to the $L_{\rm IR}$--$L'_{\rm CO(1-0)}$ relation \citep{carilli13}. ID.03 is found within the observed scatter of the relation, in the region populated by color-selected galaxies at $z>1$, and about 1 order of magnitude fainter (in terms of IR luminosity) than typical SMGs. The galaxy is quite massive and has a specific star formation rate of sSFR $\sim$ $0.15\pm0.04$ Gyr$^{-1}$ and is about 1 order of magnitude below the sSFR of an average `main--sequence' galaxy at $z\sim1.7$ \citep[e.g.][]{elbaz11}. 
Given its molecular gas mass of $\sim (9\pm2) \times 10^{10}$\,\Msun{} \citep[obtained by assuming $\alpha_{\rm CO}=3.6$ \Msun{}~(\Kkmspc)$^{-1}$; see][]{daddi10a}, we derive a star formation efficiency of SFR/$M_{\rm H2} \sim 0.4$\,Gyr$^{-1}$, or a gas depletion time of order $\sim 2$\,Gyr. 

\indent {\bf ID.08} ($\nu = 93.17\pm0.03$ GHz) and its companion ID.17 are both spatially consistent with the SMG HDF850.1. The two lines are identified as CO(5-4) and CO(6-5) at $z=5.183$. In the Appendix \ref{sec_co76} we present additional observations of this source, encompassing the CO(7-6) and \Ci{}$_{2-1}$ lines. We refer to \citet{walter12} for more details on this galaxy. 

\indent {\bf ID.17} ($\nu = 111.85\pm0.06$ GHz), together with ID.08 (see above), pins down the redshift of the SMG HDF850.1 at $z=5.183$.

\indent {\bf ID.19} ($113.45\pm0.04$ GHz) is associated with a galaxy at RA=12:36:51.61, Dec=+62:12:17.3, for which grism spectroscopic redshift $z_{\rm grism}=2.044\pm0.002$ is available through AGHAST (Weiner et al.\ 2013, in prep., see Fig.~\ref{fig_id19}). This redshift is in excellent agreement with the redshift inferred from the CO line detection in our study ($z=2.0474\pm0.0015$), under the assumption that the line is identified with CO(3-2). The available photometry of this source samples a broad range of wavelengths, up to MIPS 24 $\mu$m. The source is formally not detected in Herschel maps, although a bright, nearby foreground ($z=0.300$) galaxy prevents us from directly measuring the FIR emission in the counterpart of ID.19. From the SED fit, we infer the following properties of the counterpart of ID.19: SFR=$7.9_{-1.4}^{+3.5}$ \Msun{}\,yr$^{-1}$; $M_*=(1.9\pm0.3) \times 10^{10}$\,\Msun{}; $L_{\rm dust} = (8_{-1}^{+11}) \times 10^{10}$\,\Lsun{}; $M_{\rm dust} = (9_{-7}^{+50}) \times 10^{6}$\,\Msun{}. The galaxy is thus quite massive and has a specific star formation rate of sSFR $\approx$ 0.4 Gyr$^{-1}$, below the sSFR of an average `main--sequence' galaxy at $z\sim2.05$ \citep[e.g.][]{elbaz11}. We measure a line luminosity $L'_{\rm CO(3-2)} = (9.9\pm2.6)\times10^9$ \Kkmspc{}. Assuming a Milky-Way-like CO excitation, the corresponding CO(1-0) luminosity would be $L'_{\rm CO(1-0)} \approx3.7\times10^{10}$ \Kkmspc{}. We note that this value is significantly larger than what is observed in local spiral galaxies of comparable IR luminosity. The discrepancy might be reduced by a factor of 3.5 if the true CO excitation is higher than the MW--like excitation that we assume. By adopting an M82-like CO excitation \citep{weiss07}, we obtain $L'_{\rm CO(1-0)} \approx1.1\times10^{10}$ \Kkmspc{}, which locates ID.19 within the scatter of the observed $L_{\rm IR}$--$L'_{\rm CO(1-0)}$ relation (see Fig.~\ref{fig_co_fir}), among local ultra-luminous IR galaxies. Assuming $\alpha_{\rm CO}=3.6$ \Msun{}/(\Kkmspc)$^{-1}$, following \citet{daddi10a}, we derive the total molecular gas mass $M_{\rm H2}\sim 1.3\times10^{11}$ \Msun{}, although this estimate is significantly depends on the CO excitation and on the choice of $\alpha_{\rm CO}$.

\subsubsection{High--quality line candidates}
\begin{figure}
\includegraphics[width=0.99\columnwidth]{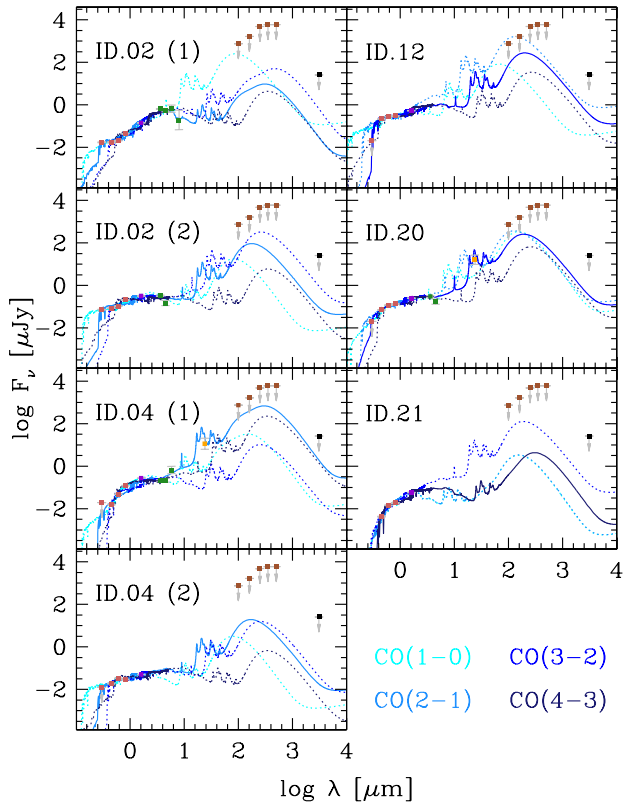}\\
\caption{Observed spectral energy distributions of the optical/NIR counterparts of other (positive) CO line candidates of our molecular line scan. Symbols are coded as in Fig.~\ref{fig_sed03}. Upper limits (3--$\sigma$) are plotted as downward arrows. For each galaxy, we fit the observed photometry assuming four different redshifts, each associated with a different possible identification of the CO line candidate discovered in our scan (from cyan to dark blue: CO(1-0), CO(2-1), CO(3-2) and CO(4-3)). Fits that are associated with our adopted $z_{\rm CO}$ are highlighted with solid lines. For ID.02 and ID.04, two possible counterparts are identified (see Tab.~\ref{tab_zlum}).}
\label{fig_sed}
\end{figure}

\indent {\bf ID.01} ($\nu=80.05\pm0.03$ GHz) is a prominent line candidate (the third most significant line, in terms of SNR$_{\rm spec}$). The line profile may be indicative of two peaks separated by $\sim300$ \kms{}, but the SNR is too low for a proper line profile characterization. The lack of a second line at the same position suggests that the source is at $z=0.440$ (if the line were identified as CO(1-0)) or $z=1.8799$ (if the line were CO(2-1)). The source has no optical/NIR counterpart, which makes the lower-$z$ interpretation unlikely.

\indent {\bf ID.02} ($\nu=82.07\pm0.02$ GHz) is at the edge of the primary beam area in one of the lowest frequency setups in the scan. It has two possible counterparts (see Fig.~\ref{fig_sed}), with $z_{\rm phot}=0.92$ and $1.12$ in \citet{soto99}. The line can be intepreted as CO(1-0) at $z=0.41$ or as CO(2-1) at $z=1.82$. The SED fit favors the latter scenario for both counterparts. The first of the two galaxies seems to be the most plausible counterpart, since it is best fitted assuming higher dust extinction in the optical bands than in the companion, implying brighter FIR emission (and therefore, likely brighter CO fluxes). An alternative scenario to explain the relatively blue color of the optical counterparts and the bright CO emission in this source (and possibly in other cases in our sample) is that the optical counterparts are acting as gravitational lenses of a background, CO--bright source. We note however that the lensing hypothesis is unlikely affecting more than a few sources in our sample (if any).

\indent {\bf ID.05} ($\nu = 89.89\pm0.04$ GHz) shows no second line at the same spatial position, suggesting $z_{\rm CO}<3.016$. The lack of any optical/NIR counterpart makes an identification as CO(1-0) ($z_{\rm CO}=0.2824$) unlikely. The two remaining options are CO(2-1) at $z=1.5648$ or CO(3-2) at $z=2.8471$. The latter is preferred due to the lack of any optical counterparts. 

\indent {\bf ID.10} ($\nu = 103.88\pm0.03$ GHz) can be identified as CO(2-1) at $z=1.219$ or CO(3-2) at $z=2.329$. In our analysis, we adopt the latter (based on the lack of optical/NIR counterparts). We note however that a few other galaxies with optical/NIR spectroscopic redshift $z\approx1.2$ are observed in the field, tentatively pointing towards a galaxy overdensity at this redshift. If that was indeed the case, the redshift assignment for ID.10 should be re--considered.

\indent {\bf ID.11} ($\nu = 108.32\pm0.03$ GHz) has no optical/NIR counterparts. Plausible interpretations are therefore CO(2--1) at $z=1.128$ or CO(3--2) at $z=2.192$ and we adopt the latter redshift, given the absence of emission at optical/NIR wavelengths.

\indent {\bf ID.12} ($\nu = 108.43\pm0.04$ GHz) is spatially consistent with a pair of galaxies detected at optical/NIR wavelengths. One of them has a redshift $z=0.9614$ \citep{barger08}, which is not consistent with any CO identification of the line detected in our 3\,mm scan. On the other hand, the second galaxy has a photometric redshift of $z=2.08$ \citep{soto99}, consistent with the identification of this line as CO(3-2) at $z=2.1891$. The SED of this galaxy suggests that this is a rather blue galaxy with $L_{\rm dust} \approx 8\times10^9$ \Lsun{}. The Herschel/PACS 100 $\mu$m observation of the field reveals a tentative detection ($\sim1$ mJy, close to the 3-$\sigma$ sensitivity limit) spatially consistent with this source. Such a flux is significantly brighter than what is predicted in our SED fits if that data point is ignored (see Fig.~\ref{fig_sed}). Since the 100 $\mu$m emission is partially blended with the much brighter BzK counterpart of ID.03, and possibly with other sources, we opt to treat this measurement as an upper limit.

\indent {\bf ID.15} ($\nu = 110.55\pm0.04$ GHz) has no counterpart (the closest galaxy in the HST/NICMOS H-band observations is $>3''$ away). We identify the line as CO(3-2) at $z=2.150$. Alternatively, the line could be CO(2-1) at $z=1.100$.

\indent {\bf ID.18} ($\nu = 112.63\pm0.04$ GHz) is the second most significant line detection in our survey. We verified that the line is likely real by splitting the data taken in this frequency setup into 4 independent blocks, and searching for the line in each of them. This test retrieves the line at the expected significance in each setup. The probability of this line being a real astrophysical source is $\sim95$ \% in the analysis by Lentati et al.~(in prep.). The line flux is $0.58\pm0.11$ \jykms{}. No other line is reported in the 3\,mm scan at this spatial position. This excludes a high-$J$ identification. On the other hand, no galaxy counterpart is observed at any optical/NIR/MIR wavelengths. The lack of counterparts likely rules out a low--$z$ interpretation (excluding anything below $z\sim1.5$, down to stellar luminosities comparable with the Large Magellanic Cloud). This suggests that the line is most likely CO(3-2) at $z=2.071$. As in all previous cases, a measurement of a second CO line is required to confirm this redshift identification.

\indent {\bf ID.20} ($113.45\pm0.05$ GHz) is associated with a galaxy at RA=12:36:50.396, Dec=+62:12:05.12. Our CO observations suggest that the source is CO(3-2) at $z=2.05$, a redshift identification that is supported by the observed photometry (see Fig.~\ref{fig_sed}). Our SED fit yields a stellar mass $M_*=(1.21\pm0.07)\times10^9$ \Msun{}, a dust luminosity of $L_{\rm dust}=(5.6\pm1.2)\times10^{10}$ \Lsun{}, and an associated SFR=$1.0\pm0.3$ \Msun{}\,yr$^{-1}$. The CO(3-2) line luminosity is $L'_{\rm CO(3-2)}=(8.8\pm2.5)\times10^9$ \Kkmspc{}. Under the same assumptions used for ID.19, we derive $L'_{\rm CO(1-0)}\approx 3.2\times10^{10}$ \Kkmspc{}.

\subsubsection{Other line candidates}

\indent {\bf ID.04} ($84.94\pm0.03$ GHz) is a low--quality line candidate in our search. It can be interpreted as CO(1-0) at $z=0.36$ or CO(2-1) at $z=1.72$. It has two possible counterparts (see Tab.~\ref{tab_zlum} and Fig.~\ref{fig_sed}). The redshift $z=1.72$ hypothesis for the second galaxy yields $L_{\rm IR}\sim7\times10^{10}$ \Lsun{}, i.e., more than 1 order of magnitude brighter IR emission than the low--redshift case. We conclude that a CO redshift at $z=1.72$ is more plausible.

\indent {\bf ID.13} ($\nu = 108.77\pm0.03$ GHz) is a low-quality line candidate without optical/NIR counterparts. If real, plausible intepretations are CO(2-1) at $z=1.120$ or, most likely, CO(3-2) at $z=2.179$.

\indent {\bf ID.14} ($\nu = 109.65\pm0.03$ GHz) lacks optical/NIR counterparts. The line is flagged as low-quality in our analysis. Plausible interpretations are CO(2-1) and CO(3-2) at $z=1.102$ and $z=2.154$, respectively. 

\indent {\bf ID.21} ($114.15\pm0.05$ GHz) is a low-quality line candidate, tentatively associated with a faint galaxy at RA=12:36:51.358, Dec=+62:12:11.89 (see Fig.~\ref{fig_sed}). Given the faintness of the source and its colors, a relatively high-$z$ scenario ($z\sim2-3$) is favored. The lack of a second CO detection in our scan would point towards the $z=2.03$ scenario, although the signal--to--noise of our CO detection is modest, and we cannot rule out the $z=3.04$ case, which is also acceptable given the available optical/NIR photometry.

%

%

\section{Comparison with expectations}\label{sec_models}

In the following analysis, we only consider the `secure' and `high--quality' line candidates.

\begin{figure}
\includegraphics[width=0.99\columnwidth]{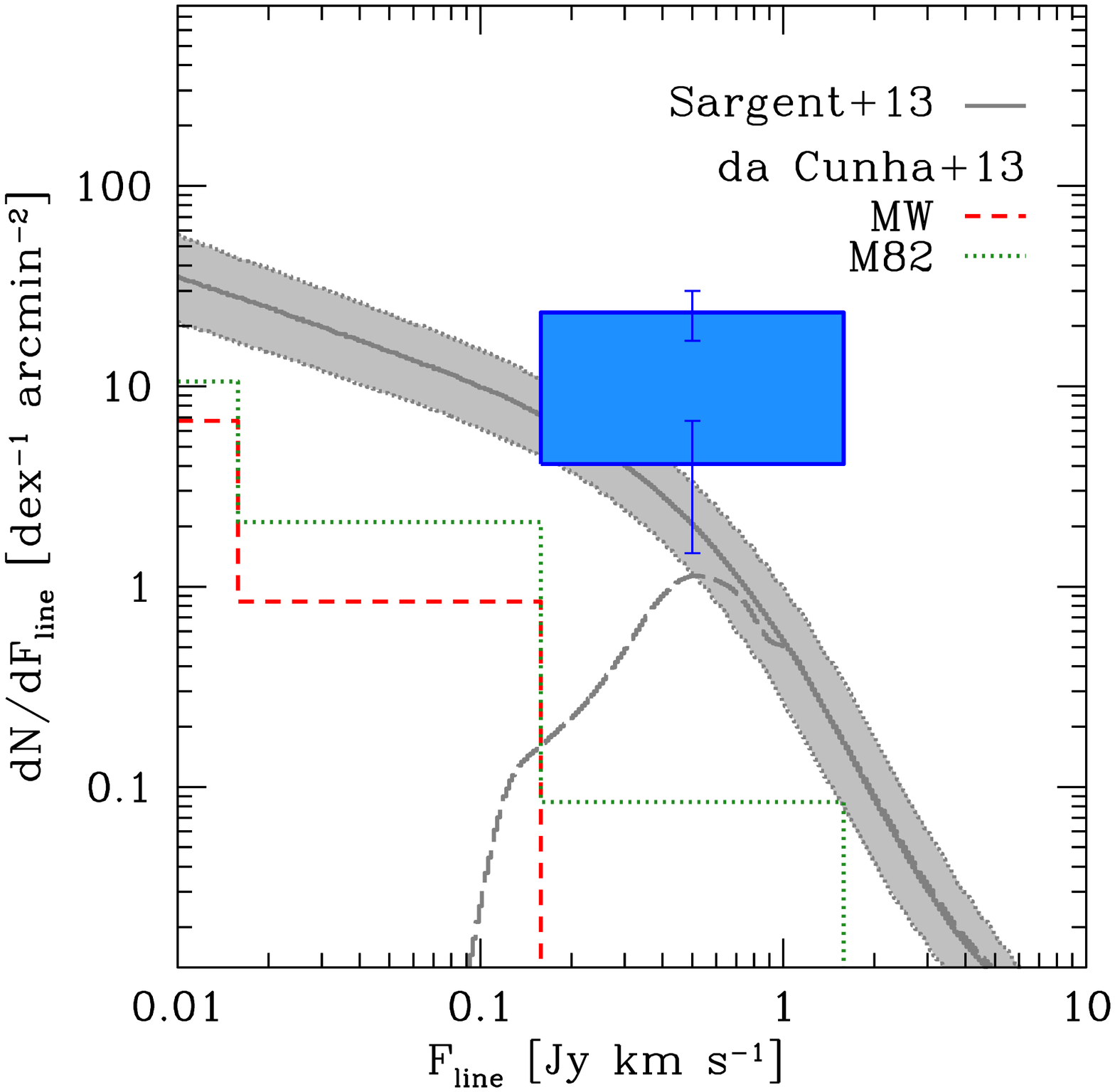}\\
\caption{Comparison between the observed number of CO line candidates (shaded box), and the expectations from the empirical predictions presented in \citet{sargent13} and \citet{dacunha13} (in the case of Milky Way-like CO excitation and in the case of M82-like excitation). The grey shading shows the 1-$\sigma$ uncertainties in the predictions by \citet{sargent13}. The long-dashed line shows the expected {\em observed} distribution, based on the predictions by \citet{sargent13} multiplied by the detection fration discussed in \S\ref{sec_complete}. The shaded area spans a 1-dex CO flux bin (encompassing all the line candidates in our study). The bottom side is set by the two secure detections ID.03 and ID.19 (HDF850.1 is not included here, since it is not representative of the general population of galaxies in the field). The upper side of the box is set by the total number of high--quality line candidates. In the area normalization, we took into account the actual beam size at the frequency of each line candidate. Error bars show poissonian uncertainties in the observed line flux distribution. }
\label{fig_mark}
\end{figure}

In Fig.~\ref{fig_mark} we compare the total number of line candidates in our scan with empirical predictions. From the list of all the positive line candidates in Tab.\ref{tab_lines}, we have omitted the line candidates ID.08 and ID.17, which are associated with HDF850.1, as they are not representative of the general field population of galaxies (our pointing centre was chosen such to include HDF850.1). We are therefore left with 2 secure line detections, plus 9 high--quality (positive) line candidates. The contribution of each line is normalized by the actual size of the primary beam at the frequency of each line. All the lines are then grouped into a single, 1-dex wide flux bin.

We compare the observed flux distribution of our line candidates with the expected distributions of low-$J$ CO transitions based on the predictions by \citet{sargent12,sargent13} and \citet{dacunha13}. \citet{sargent12,sargent13} derived empirical predictions for the CO and IR luminosity functions of galaxies and for their evolution as a function of cosmic time. The framework is a description of star formation as a two--mode phenomenon. This divides galaxies into a ``main sequence'' and a ``starburst'' population \citep[see also][]{elbaz11}. The main driver of this distinction is the specific SFR (sSFR): Star formation efficiencies for main sequence galaxies are based on the observed Schmidt--Kennicutt law; starbursts deviate from this law due to an enhancement in their SFRs. The CO--to--H$_2$ conversion factor $\alpha_{\rm CO}$ is assumed to depend on the metallicity ($Z$), scaling as $\sim Z^{-1}$ for main sequence galaxies, while for starbursts $\alpha_{\rm CO}$ is strongly sensitive to the SFR-enhancement. Once folded into the evolution of the stellar mass function of star-forming galaxies and the evolution of the sSFR, this approach infers IR luminosity function of main sequence and starbursting galaxies, and thus CO(1-0) luminosity functions. These are then expanded towards higher $J$ by assuming the CO excitation ladder of BzK 21000 \citep{dannerbauer09,daddi10a,aravena10,casey11} for main-sequence galaxies at $z>1$, and of GN20 \citep{daddi09,carilli11,bothwell13} for starbursts. The resulting distributions are dominated by $J<4$ transitions for CO fluxes $\gsim0.1$ \jykms{}.

\citet{dacunha13}, on the other hand, fitted the spectral energy distribution of galaxies in the Hubble Ultra Deep Field using \textsf{magphys} \citep{dacunha08}\footnote{The predictions are done for the Hubble Ultra Deep Field (UDF) which is not the HDF--N studied here. We chose the UDF comparison as the available multi--wavelength data needed for the predictions is deeper and more complete for the UDF. We do not expect significant differences in galaxy properties between the HDF--N studied here and the UDF.}. By assuming energy balance between the absorbed--UV radiation from young stars and the reprocessed light from dust, this method predicts dust luminosities based on observed optical/NIR/MIR photometry. The IR luminosities predicted with these two methods are then transformed into $L'_{\rm CO}$ via the empirical relation between CO(1-0) and dust luminosity \citep{daddi08,genzel10,carilli13}. Finally, \citet{dacunha13} derive fluxes of higher-$J$ transitions by assuming two extreme cases, a low-CO-excitation scenario, with a Milky Way-like CO ladder, and a high-CO-excitation scenario, with a CO ladder resembling the center of the starbursting galaxy M82. The CO line energy distributions in these two cases are taken from \citet{weiss07}.

The observed distribution resulting from our blind CO search suggests an excess of sources with respect to the predictions, in particular if we consider all the 11  high--quality line candidates. We stress that a fraction of them is likely not real. We discuss the implications of our observations regarding the CO luminosity functions and the  molecular gas density of the universe as a function of redshift in W13.

\begin{table}
\caption{\rm Catalogue of the (4 positive, 1 negative) sources detected within the primary beam of our observations in the collapsed 3\,mm continuum. Fluxes and uncertainties are corrected for primary beam attenuation.} \label{tab_cont_sources}
\begin{center}
\begin{tabular}{ccccc}
\hline
ID  & Alt.name  & RA          & Dec         & Flux \\
    &           & (J2000.0)   & (J2000.0)   & [$\mu$Jy] \\
 (1)& (2)       & (3)         & (4)         & (5)  \\
\hline
C1  & HDF850.1  & 12:36:51.94 & +62:12:26.0 & $56\pm10$ \\     
C2  &           & 12:36:48.50 & +62:12:32.4 & $38\pm11$ \\     
C3  &           & 12:36:47.57 & +62:12:35.2 & $46\pm15$ \\     
C4  &           & 12:36:47.99 & +62:12:17.5 & $39\pm12$ \\     
C5  &           & 12:36:50.63 & +62:12:42.4 & $-47\pm12$ \\        
\hline
\end{tabular}
\end{center}
\end{table}

\section{The continuum image}\label{sec_cont}
\begin{figure*}
\includegraphics[angle=-90, width=0.99\textwidth]{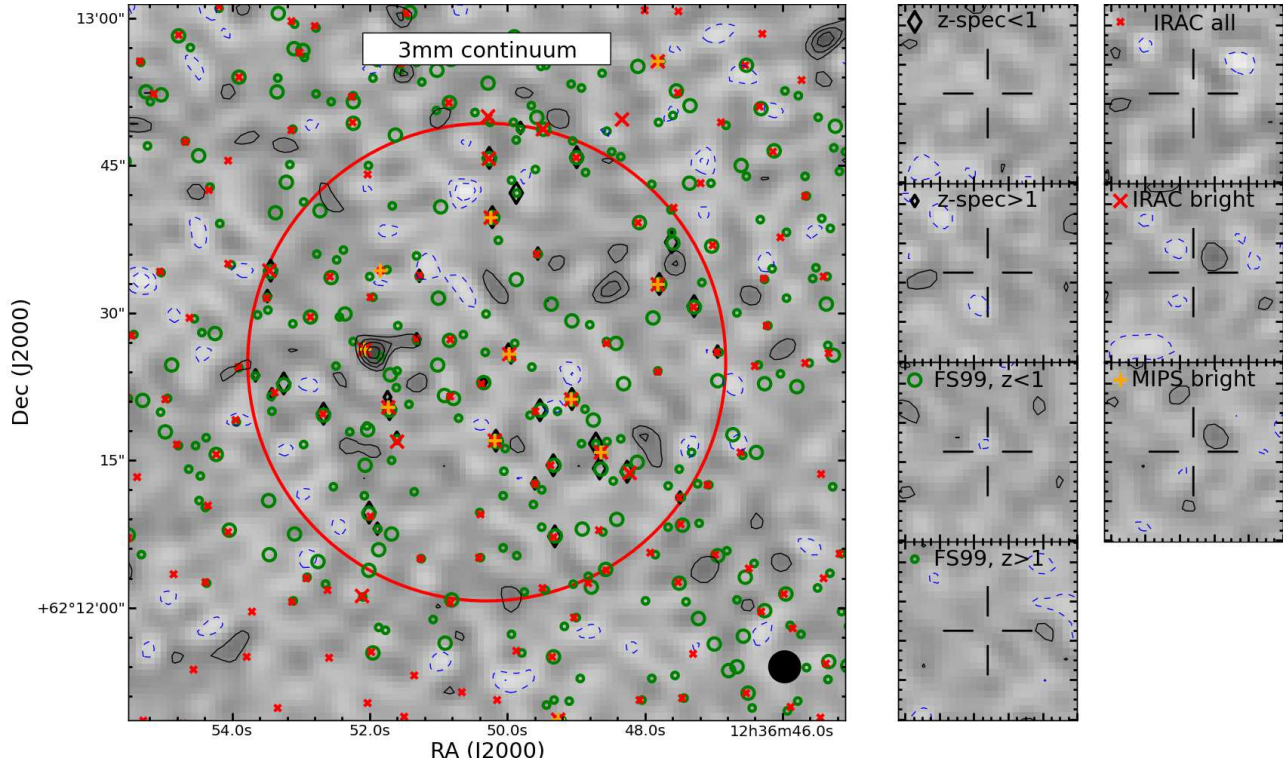}\\
\caption{{\em Left:} 3\,mm continuum image of the region of the HDF--N covered in our scan. Solid black / dashed blue contours show positive / negative 2, 3, 4, \ldots $\sigma$ contours (1\,$\sigma$=8.6 $\mu$Jy\,beam$^{-1}$). The big solid red circle shows the FWHM of the primary beam size at 97.25\,GHz (i.e., the central frequency of our scan). Black diamonds mark the position of galaxies with spectroscopic redshifts (big / small symbols refer to objects at $z_{\rm spec} > 1$ / $< 1$). Green circles label optically-selected galaxies from \citet{soto99} (FS99, bigger symbols highlight galaxies with photometric redshift $z_{\rm phot}<1$). Red crosses mark IRAC-detected sources (bigger symbols identify the ``IRAC-bright'' sample, i.e., sources with IRAC\,$3.6$ $\mu$m fluxes brighter than 3.5\,$\mu$Jy). Orange ``+'' symbols show MIPS-selected galaxies, i.e., sources with MIPS 24 $\mu$m fluxes brighter than 20\,$\mu$Jy. The beamsize of the observations is indicated in the bottom right corner. {\em Right:} For each subsample of galaxies, we show the postage stamp of the stacked emission. Each panel is $18''\times18''$ wide (i.e., same scale as in the main panel). Only sources within the primary beam are considered in the stacking analysis. No detection is reported in any of the samples down to stacked continuum limits of only a few $\mu$Jy (Tab.~\ref{tab_stack}).}
\label{fig_cont}
\end{figure*}

By collapsing our molecular line scan in frequency space we obtained a 3\,mm continuum map with an rms of $8.6$ $\mu$Jy\,beam$^{-1}$, making this by far the deepest map at this frequency to date (see Fig.~\ref{fig_cont}). The image has been cleaned down to the 2\,$\sigma$ level. We report four sources brighter than $>$3\,$\sigma$, all consistent with being spatially unresolved. Coordinates and fluxes are reported in Tab.~\ref{tab_cont_sources}. The brightest detection (ID.C1) is associated with the SMG HDF850.1, and it is spatially consistent with the lines ID.08 and ID.17 in Tab.~\ref{tab_lines}. ID.C4 is marginally consistent with the line candidate ID.01 (the offset is $2.7''$). For the remaining two continuum sources (ID.C2 and ID.C3), we do not report any significant line association in our 3\,mm scan. They are likely spurious, as suggested by the detection of an equally-significant ($\gsim$3\,$\sigma$) negative peak (dubbed ID.C5).

The deep continuum map allows us to constrain the observed SED of known galaxies at an observed wavelength of 3\,mm. We stack the 3\,mm continuum image at the coordinates of optical/NIR/MIR-selected galaxies. The stacks are performed both in the (U,V) and sky plane, focusing on the following samples: 1) sources with spectroscopic redshift from optical/NIR studies \citep[][plus additional 8 unpublished slit redshifts and 23 grism redshifts, see W13]{bundy09}, split into galaxies above and below $z=1$ (in order to have similar sub-sample sizes at low and high $z$). As discussed in W13, these samples are highly complete up to $z\sim3$ for $m_{\rm H}<24$ mag (roughly corresponding to $M_*=7\times10^9$ \Msun{} at $z=2-3$); 2) optically-selected galaxies from the compilation in \citet{soto99}, split again according to their photometric redshift (below or beyond $z_{\rm phot}=1$); 3) IRAC 3.6$\mu$m-detected sources from the Spitzer Legacy Project on GOODS (PI: M.~Dickinson); 4) a sub-sample of the IRAC list, selected by requiring $3.6$ $\mu$m fluxes brighter than 3.5 $\mu$Jy; 5) MIPS-selected galaxies, i.e., sources with MIPS 24 $\mu$m fluxes brighter than 20 $\mu$Jy. From Fig.~\ref{fig_cont}, it is apparent that there is substantial overlap among some galaxy samples, especially if only the brightest sources are considered. We note that each sample consists of $<$100 sources, i.e., the sample sizes are all significantly lower than the number of the resolution elements in the map.

The stacking technique is similar with the one presented in \citet{decarli13_aless}, and works as follows: We crop the input catalogues to consider only sources within the primary beam of our observations. As for the reference frequency (needed to compute the size of the primary beam), we refer to the central frequency of the scan (97.25 GHz), which implies a fiducial diameter of the primary beam of $48.6''$. For each source, we compute its offset with respect to the pointing center. These offset values are used to correct for primary beam attenuation and to compute weights, so that a source lying at the primary beam radius is scaled up by a factor 2 in flux and is assigned a weight of 0.25 (i.e., the squared value of the primary beam attenuation). In the (U,V) stacking, we shift the sources into the pointing center by applying phase shifts, and then we stack amplitudes. In the image plane stacking, we cut out postage stamps of the sources from the continuum map, realign and combine. In order to quantify the contribution of bright outliers we apply a jack-knife analysis: for each sample of $N$ galaxies, we perform $N$ stacks of $N-1$ sources, by removing a different source each time. Finally, we quantify the uncertainties by stacking on random coordinates within the primary beam. The stacking uncertainties are estimated as the standard deviations of the stacked fluxes over 50 realizations. 

No detection is reported in the stacked 3\,mm continuum image of any of the samples considered in our stacking analysis. Tab.~\ref{tab_stack} summarizes the results of this stacking experiment.  We achieve 1-$\sigma$ stack sensitivities of a few $\mu$Jy in all samples. We compute the associated limits on the dust continuum emission by assuming a fiducial redshift $z=1$ for all the samples, and shifting the M\,51 and M\,82 templates by \citet{silva98} in order to match the 3-$\sigma$ limits from the stacks at the central frequency of our scan (97.25 GHz, our results are practically unchanged for different assumptions of the fiducial redshift, for any $z\gsim1$). The M\,51 and M\,82 templates are adopted to represent a typical spiral galaxy and a starbursting galaxy, respectively. 

\begin{figure}
\includegraphics[width=0.99\columnwidth]{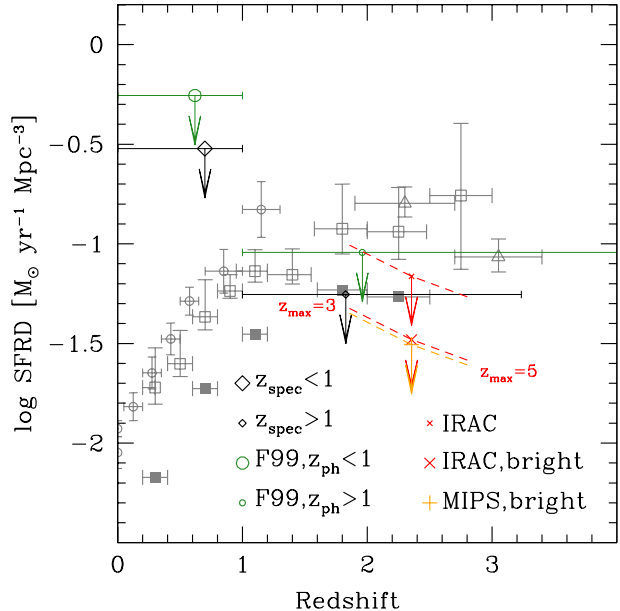}\\
\caption{Constraints on the cosmic star formation rate density resulting from our stacking analysis of the 3\,mm continuum image presented in this work, assuming an M51-like dust template to convert observed 3\,mm fluxes into IR luminosities. Symbols are the same as in Fig.~\ref{fig_cont}. The 3-$\sigma$ limits quoted in Tab.~\ref{tab_stack} have been divided by the comoving volume covered by each galaxy sample, and scaled up to take into account the effects of CMB, following \citet{dacunha13b}. For galaxies with redshift information, the position of the points along the x-axis is set by the average redshift of each sample. As for the samples of galaxies without redshift information, we assume $z_{\rm min}=0$ and $z_{\rm max}=4$, and place our constraint at the volume-averaged redshift. Dashed lines show how our limits would change if $z_{\rm max}$ changed from $z=3$ to $z=5$. 
For a comparison, we plot IR/radio-based SFR density estimates from the literature: empty gray squares from \citet{karim11}, circles from \citet{rujopakarn10} and triangles from \citet{reddy08}, homogenized to a \citet{chabrier03} initial mass function. Filled squares show the results from \citet{karim11}, once we consider only galaxies with $M_*>10^{10}$ \Msun{}. We note that our constraints on SFRD would shift up-wards if a template with higher $T_{\rm dust}$ was adopted. E.g., using M82 as a template would imply $\sim8\times$ higher SFRDs for the same 3\,mm fluxes.}
\label{fig_madau}
\end{figure}

Our 3\,mm continuum observations sample the Rayleigh--Jeans tail of the dust emission. Therefore, for a given 3\,mm continuum flux, the inferred luminosity (as reported in Tab.~\ref{tab_stack}) is a factor $\sim7$ brighter using the M\,82 template compared to the M\,51 case, due to the higher dust temperature. Our 3\,$\sigma$ IR luminosity limits are in the range $(5 - 20)\times 10^{10}$ \Lsun{} for M51--like galaxies, and $(4-13) \times 10^{11}$ \Lsun{} for M82--like dust emission. We convert these limits into constraints on the associated SFR by assuming SFR/[\Msun{}\,yr$^{-1}$] = $1.3 \times L_{\rm IR}$/[$10^{10}$ \Lsun] \citep{genzel10}. The resulting SFR 3-$\sigma$ limits in our galaxy samples range between 7 and 23 \Msun{}~yr$^{-1}$ in the spiral galaxy case and between 50 and 170 \Msun{}~yr$^{-1}$ in the starburst case. We compute the contribution of each sample to the cosmic SFR densities (SFRD, see Fig.~\ref{fig_madau}) in the case of an M51--like dust template. In order to compute the sampled cosmic volume in each galaxy set, we proceed as follows: 1) For galaxy samples with spectroscopic or photometric $z<1$, we adopt $z_{\rm min}=0$, $z_{\rm max}=1$; 2) For the $z>1$ samples, the upper end of the cosmic volume is fixed by the highest-$z$ galaxy in the sample; 3) For galaxy samples without available redshift information, we assume a broad redshift window ranging between $z_{\rm min}$=0 and $z_{\rm max}$=3--5. Following \citet{dacunha13b}, we correct for the cosmic microwave background (CMB), which acts as an observing background and sets a minimum dust temperature. The net effect is that only a fraction of the intrinsic flux of the Rayleigh-Jeans tail of the dust SED is observed. At 3\,mm, the correction is $\lsim5$\% at $z<1$, and up to $\sim50$\% at $z=3$ for a relatively low dust temperature (as in the M51 case) \citep{dacunha13b}. 

We find that at $z<1$, the limits on the cosmic SFRD set by 3\,mm observations are not stringent, once compared with the observed SFRD evolution \citep[we here compare with the radio/IR-based results by][]{reddy08,rujopakarn10,karim11}. This is partially due to the fact that the cosmic volume probed in our observations at $z<1$ is small. The $z>1$ spectroscopic and photometric samples account for up to about 50 and 65\% of the total cosmic SFRD, respectively. We estimate a similar contribution for IRAC--detected galaxies (although with larger uncertainties, due to the assumptions on the sampled cosmic volume). Half of this contribution comes from IRAC--bright and MIPS--bright sources. If we adopt an M82--like dust template, our SFRD limits are shifted up by a factor $\sim8$. Our results suggest that, at $1<z\lsim3$, optical/MIR bright galaxies with $M_*\gsim10^{10}$ \Msun{} contribute only to a minor fraction of the SFRD at those redshifts, unless their dust SED is better described by starburst template (M82--like, with $T_{\rm dust}\approx40$ K) than by local spiral galaxies (M51--like, with $T_{\rm dust}\approx15$ K). \citet{karim11} studied the cosmic SFRD as a function of both redshift and stellar masses. Their results show that star-forming galaxies with $M_*>10^{10}$ \Msun{} contribute to about 50\% of the total SFRD at any redshift between $z$=0 and $z\sim 2.5$, in broad agreement with our findings.

\begin{table}
\caption{\rm Limits on the 3\,mm continuum from the stacking analysis. The luminosity limits are computed by assuming a fiducial $z=1$, the M51/M82 templates by \citet{silva98}, and a 3-$\sigma$ detection. Associated star formation rates are derived assuming SFR/[\Msun{}\,yr$^{-1}$] = $1.3 \times L_{\rm IR}$/[$10^{10}$ \Lsun] \citep{genzel10}.} \label{tab_stack}
\begin{center}
\begin{tabular}{ccccccccc}
\hline
Sample           & N   & RMS & $z_{\rm min}$ & $z_{\rm max}$   & \multicolumn{2}{c}{log $L_{\rm IR}$} & \multicolumn{2}{c}{SFR}\\
                 &     &  & & & M51 & M82 & M51 & M82 \\
                 &     & [$\mu$Jy] & & & \multicolumn{2}{c}{[\Lsun{}]} & \multicolumn{2}{c}{[\Msun\,yr$^{-1}$]}\\
 (1)             & (2) & (3) & (4)   & (5)   & (6) & (7) & (8) & (9) \\
\hline
$z_{\rm spec}<1$ &  20 & 3.9 & 0   & 1       & 11.04 & 11.92 & 14  & 110 \\
$z_{\rm spec}>1$ &  17 & 4.0 & 1   & 3.23    & 11.05 & 11.93 & 15  & 110 \\
F99, $z<1$       &  48 & 3.0 & 0   & 1       & 10.93 & 11.81 & 11  &  80 \\
F99, $z>1$       &  88 & 2.5 & 1   & 5.64    & 10.85 & 11.73 &  9  &  70 \\
IRAC             &  43 & 2.9 & 0   & $\sim4$ & 10.91 & 11.79 & 11  &  80 \\
IRAC bright      &  12 & 5.0 & 0   & $\sim4$ & 11.15 & 12.03 & 18  & 140 \\
MIPS bright      &   9 & 6.3 & 0   & $\sim4$ & 11.25 & 12.13 & 23  & 170 \\
\hline
\end{tabular}
\end{center}
\end{table}

\section{Conclusions}\label{sec_conclusions}

We have obtained the first frequency scan of the 3\,mm window (79.7--114.8\,GHz) in the Hubble Deep Field North (HDF--N) using the Plateau de Bure Interferometer. This molecular line scan constitutes the first blind survey of CO emission in `normal' galaxies (i.e., non--SMGs) at high $z$. Our main observational results can be summarized as follows:
\begin{itemize}

\item[{\em i-}] We achieved approximately uniform sensitivity ($\sim0.3$ mJy\,beam$^{-1}$ per 90 \kms{} channel) over the entire scan, which allows us to probe the CO line luminosities down to $(4-8)\times10^{9}$ \Kkmspc{} at any $z>1$.

\item[{\em ii-}] Using different search algorithms we have identified 17 positive line candidates (and 4 negative ones). Typical fluxes are in the range 0.16--0.56 \jykms{}, while line widths are between 110 and 500 \kms{}.  We estimate that our completeness is $\sim66$\% for the faintest targets, and close to 100\% for the brightest. The rate of false positives is about $2$ in the whole scan.

\item[{\em iii-}] Two line detections (ID.08 and ID.17) are spatially consistent with the SMG HDF850.1 at $z=5.183$ \citep{walter12}. In the Appendix of this paper we present new 2\,mm observations encompassing the CO(7-6) and \Ci{}$_{2-1}$ lines of this source.

\item[{\em iv-}] No second line is reported at the position of any of the other line candidates, suggesting that, if real, the candidate lines belong to galaxies at intermediate to low redshift ($z\lsim3$). 

\item[{\em v-}] Besides HDF850.1, two line candidates (ID.03 and ID.19) are considered `secure', being independently confirmed by corollary data. ID.03 is the brightest line in our scan. It is associated with a FIR--bright BzK galaxy. Our follow-up observations with PdBI at 2\,mm and with JVLA in Q band allow us to pin down the redshift of this source $z=1.784$, in excellent agreement with the photometric redshift and its SED. This galaxy lies within the scatter of the `main sequence' relation of galaxies. For this source, we report a higher CO excitation than what has been observed in other BzK galaxies to date. The other secure line, ID.19, has also an optical/NIR counterpart, and fulfills the classification of a BzK galaxy. This source has an available (grism) redshift, consistent with the one inferred from our CO detection, at $z=2.048$. Among the remaining line candidates, nine (ID.01, 02, 05, 10, 11, 12, 15, 18, 20) are considered `high--quality'.

\item[{\em vi-}] We present a detailed study of optical/NIR counterparts of our line candidates. We found one or more optical/NIR galaxies that are located within the positional uncertainties of 7 line candidates (ID.02, 03, 04, 12, 19, 20, 21). We model the optical/NIR/MIR photometry of these tentative counterparts assuming various CO identifications of the line candidates, and we use the goodness of the fit as a criterion to discriminate among different redshift scenarios, when no corollary spectroscopic information is available. We note that some high--quality CO line candidates in our survey (e.g., ID.18) lack of any counterpart at other wavelengths, despite the exquisite depth of the available optical/NIR/MIR data.

\item[{\em vii-}] We report a possible excess in the fluxes of our line candidates with respect to the predictions based on the evolution of galaxy luminosity functions \citep{sargent12,sargent13} as well as extrapolation from optical/NIR emissions \citep{dacunha13}. However, we caution that this comparison includes all the `secure' and `high--quality' line candidates, although some of them may not be real (as they still need to be confirmed through follow--up observations of higher--$J$ lines). In an accompanying paper (W13), we use our measurements to constrain the CO luminosity function and the evolution of the H$_2$ mass density as a function of lookback time.

\item[{\em viii-}] By collapsing our data cube along the frequency axis, we obtain the deepest 3\,mm continuum image available to date. We clearly detect HDF850.1, and three additional continuum candidates at $\sim3$-$\sigma$ level. Only one of these fainter sources show a tentative association with a line candidate (ID.01). Neither HDF850.1 nor any of the three other continuum candidates is detected in any optical or IR survey.

\item[{\em ix-}] Using various samples of galaxies selected based on their optical or MIR emission, we place sensitive constraints on the dust continuum emission of galaxies at 3\,mm. Our results suggest that, at $1\lsim z\lsim3$, optical and MIR bright galaxies with $M_*\gsim10^{10}$ \Msun{} contribute only to a fraction ($<50$\%) of the SFRD at those redshifts, unless their dust SED is better described by starburst template (M82--like, with $T_{\rm dust}\approx40$ K) than by local spiral galaxies (M51--like, with $T_{\rm dust}\approx15$ K).
\end{itemize}

This study proves the feasibility of sensitive blind molecular line searches in distant galaxies. Parallel endeavors in the radio window (e.g., using JVLA to target low-$J$ CO transitions of high-$z$ galaxies) and at $\sim1$mm wavelengths (e.g., in the near future capitalizing on the upgraded capabilities of NOEMA/PdBI and ALMA), will provide us with a new, complete description of the molecular gas properties of star--forming galaxies through cosmic times.

\section*{Acknowledgments}
We thank the referee, S. Serjeant, for his useful comments that improved the quality of our paper. This work is based on observations carried out with the IRAM Plateau de Bure Interferometer. IRAM is supported by INSU/CNRS (France), MPG (Germany) and IGN (Spain). This research made use of Astropy, a community-developed core Python package for Astronomy \citep{astropy}. Support for RD was provided by the DFG priority program 1573 ``The physics of the interstellar medium''. FW, DR and EdC acknowledge the Aspen Center for Physics where parts of this manuscript were written.

\appendix

\section{2mm observations of HDF850.1}\label{sec_co76}

We here present PdBI observations at 2\,mm wavelength of the CO(7-6) and \Ci{}$_{2-1}$ transitions (rest-frame frequencies: 806.652 GHz and 809.342 GHz, respectively) in HDF850.1 that have been obtained as a follow--up of the molecular line scan discussed in this paper. These new observations were carried out in 4 tracks between December 3, and December 11, 2011. The receiver frequency was tuned to 130.680 GHz, thus encompassing CO(7-6) and \Ci$_{2-1}$ at $z$=5.183. At this frequency, the primary beam of PdBI antennas is $36.2''$ (FWHM). The pointing center of the observations was at RA=12:36:51.90, Dec=+62:12:25.7 (J2000.0). The array was in compact configuration (5 or 6 antennas), with baselines ranging between 17.7 m and 97.6 m. Data were reduced and analysed using GILDAS, as described in \S\ref{sec_observations}. We binned the cube into 90 \kms{} channels. The resolution element is $4.0''\times3.8''$. The final cube has 7018 visibilities, corresponding to 5.85 hr on source (6--antenna equivalent). The noise per 90 \kms{} channel is 0.60 mJy\,beam$^{-1}$.

\begin{figure}
\includegraphics[width=0.54\columnwidth]{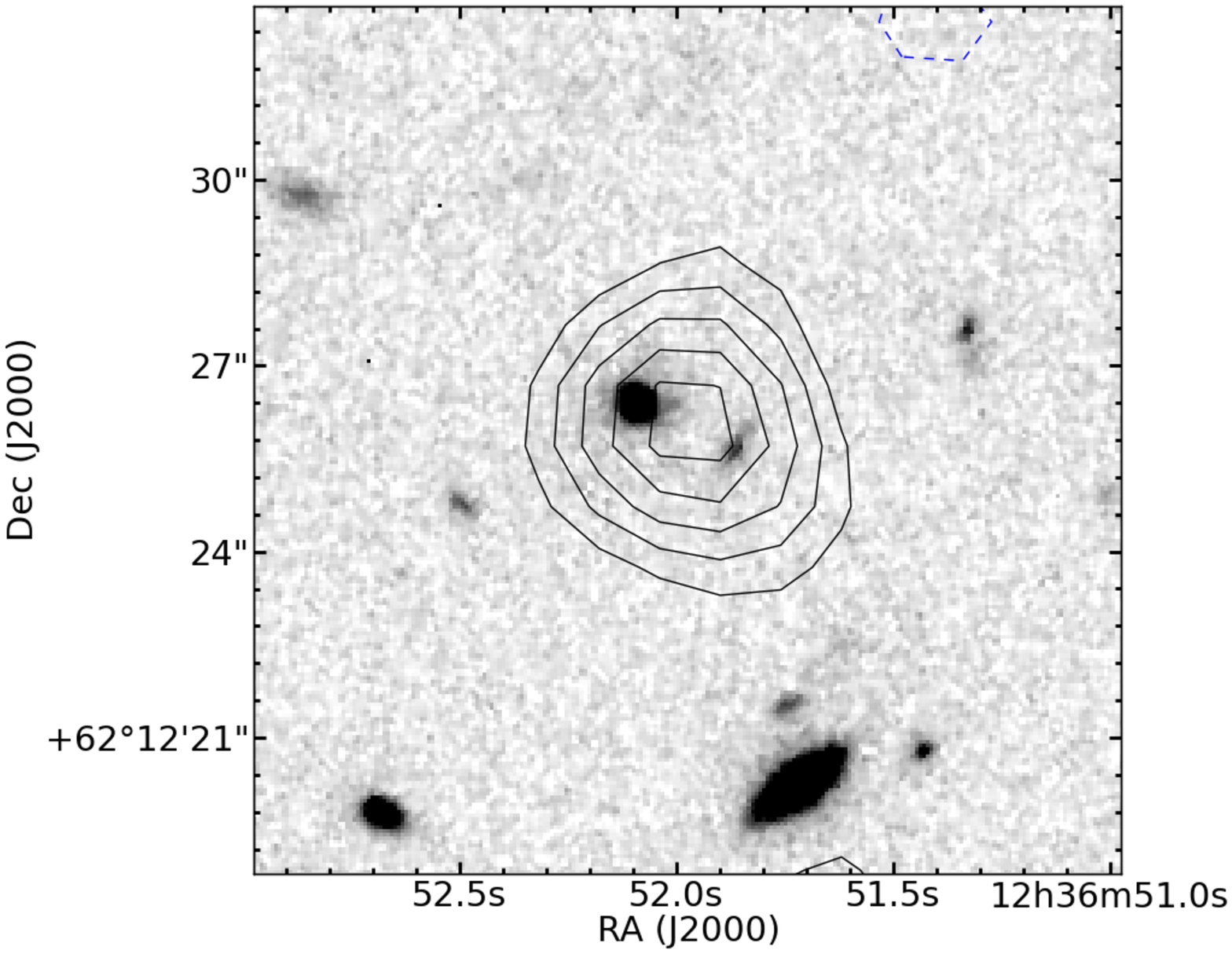}
\includegraphics[width=0.45\columnwidth]{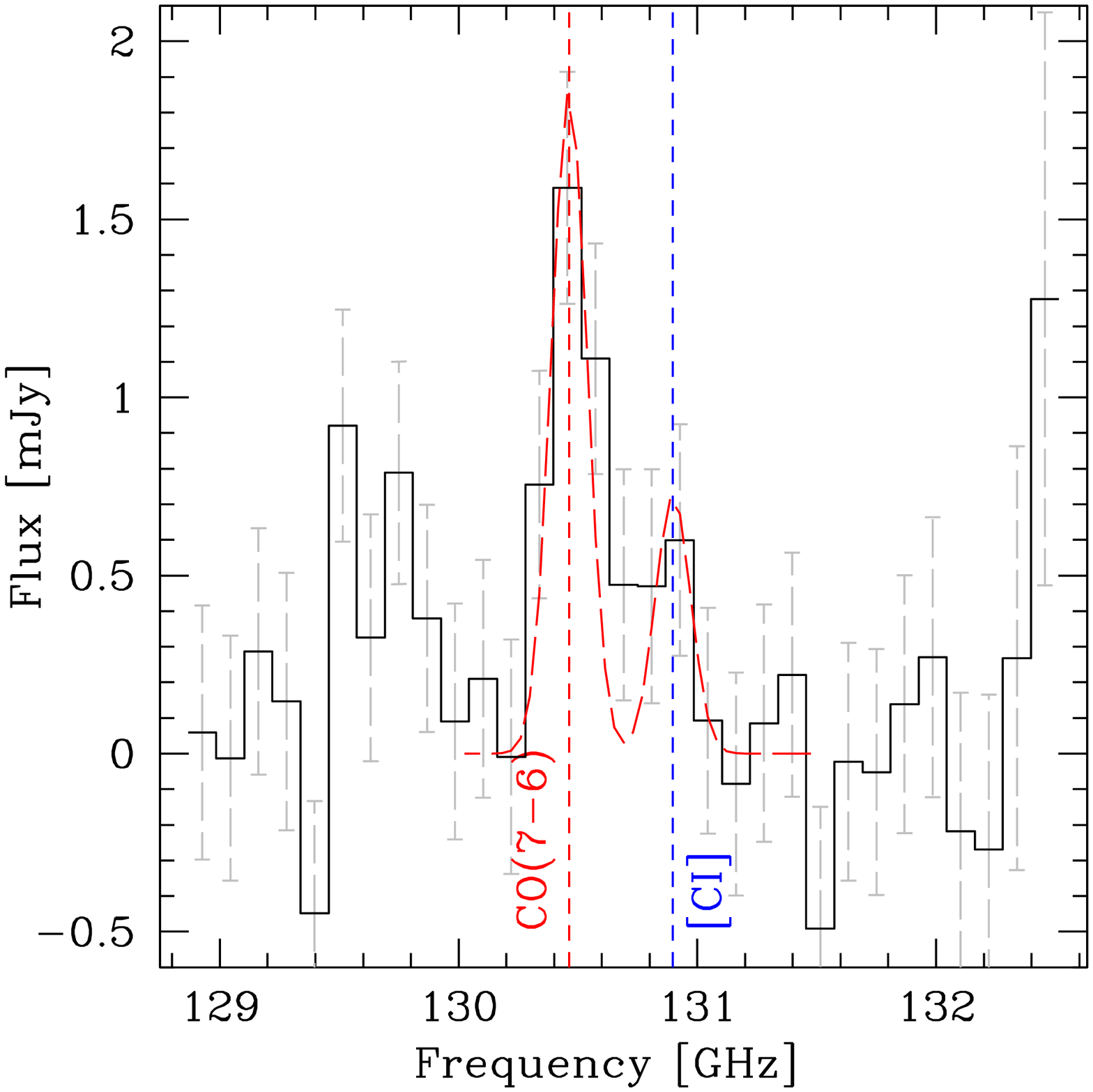}\\
\caption{Line map and spectrum of the CO(7-6)+\Ci{} system of HDF850.1 at z=5.183. {\em Left:} The line map is integrated over 1530 \kms{}, encompassing both CO(7-6) and \Ci{}. The gray scale is the HST/WFC3 F160W image of the field from the CANDELS survey \citep{grogin11,koekemoer11}. Black solid / blue dashed contours show the 2, 3, 4, \ldots $\sigma$ positive / negative isophotes in the CO map. The box is $14''\times14''$ wide. {\em Right:} The observed CO(7-6)+\Ci{} line profile is plotted as a solid histogram, rebinned to 270 \kms{} channel width. We fit the line profiles with two gaussians (red, dashed lines) fixed in redshift and width at the parameters obtained from the CO(5-4) and CO(6-5) transitions \citep[line candidates ID.08 and ID.17 in this work; see][]{walter12}.}
\label{fig_co76}
\end{figure}

The resulting spectrum of the CO(7-6) and \Ci{}$_{2-1}$ transitions extracted on the position of HDF\,850.1 is presented in Fig.~\ref{fig_co76}. CO(7-6) is detected at $6$-$\sigma$ level, and there is a 2-$\sigma$ peak at the frequency of \Ci{}. We fit these two peaks with two Gaussians with fixed peak frequency and width, as derived from the lower-$J$ CO transitions ($z=5.183$, FWHM$_{\rm CO}$=400 \kms{}). We obtain $S_{\rm CO(7-6)}=1.86\pm0.30$ mJy and $S_{\rm [CI]}=0.73\pm0.30$ mJy, yielding total line fluxes of $0.35\pm0.05$ \jykms{} and $0.14\pm0.05$ \jykms{}, and line luminosities $L=(2.4\pm0.4)\times 10^{8}$ \Lsun{}, $L'=(1.4\pm0.2)\times 10^{10}$ \Kkmspc{}, and $L=(5.7\pm2.3)\times 10^{7}$ \Lsun{}, $L'=(1.5\pm0.6)\times 10^{10}$ \Kkmspc{}, for CO(7-6) and \Ci{}$_{2-1}$ respectively. \citet{walter12} based on the CO(2-1), CO(5-4) and CO(6-5) lines infer a CO(1-0) line luminosity of $L'=4.3 \times 10^{10}$ \Kkmspc{}. This implies a $L'$(CO(7-6))/$L'$(CO(1-0)) = 0.33, which, within the uncertainties, is thus is broadly consistent with other high--redshift FIR--bright galaxies \citep{carilli13,riechers13}.

\label{lastpage}


\begin{thebibliography}{99}
\bibitem[\protect\citeauthoryear{Aravena et al.}{2010}]{aravena10} Aravena M., Carilli C., Daddi E., Wagg J., Walter F., Riechers D., Dannerbauer H., Morrison G.E., Stern D., Krips M., 2010, ApJ, 718, 177
\bibitem[\protect\citeauthoryear{the Astropy collaboration}{2013}]{astropy} The Astropy collaboration:  Robitaille T.P., Tollerud E.J., Greenfield P., Droettboom M., Bray E., Aldcroft T., Davis M., Ginsburg A., et al., 2013, A\&A, in press (arXiv:1307.6212)
\bibitem[\protect\citeauthoryear{Barger et al.}{2008}]{barger08} Barger A.J., Cowie L.L., Wang W.-H., 2008, ApJ, 689, 687
\bibitem[\protect\citeauthoryear{Bothwell et al.}{2013}]{bothwell13} Bothwell M.S., Smail I., Chapman S.C., Genzel R., Ivison R.J., Tacconi L.J., Alaghband-Zadeh S., Bertoldi F., et al. 2013, MNRAS, 429, 3047
\bibitem[\protect\citeauthoryear{Bundy et al.}{2009}]{bundy09} Bundy K., Ellis R.S., Conselice C.J., 2009, yCat, 7246, 0
\bibitem[\protect\citeauthoryear{Carilli et al.}{2011}]{carilli11} Carilli C.L., Hodge J., Walter F., Riechers D., Daddi E., Dannerbauer H., Morrison G.E., 2011, ApJ, 739, L33
\bibitem[\protect\citeauthoryear{Carilli \& Walter}{2013}]{carilli13} Carilli C.L. \& Walter F., 2013 (arXiv:1301.0371)
\bibitem[\protect\citeauthoryear{Casey et al.}{2011}]{casey11} Casey C.M., Chapman S.C., Neri R., Bertoldi F., Smail I., Coppin K., Greve T.R., Bothwell M.S., et al. 2011, MNRAS, 415, 2723
\bibitem[\protect\citeauthoryear{Chabrier}{2003}]{chabrier03} Chabrier G., 2003, PASP, 115, 763
\bibitem[\protect\citeauthoryear{da Cunha et al.}{2008}]{dacunha08} da Cunha E., Charlot S., Elbaz D., 2008, MNRAS, 388, 1595
\bibitem[\protect\citeauthoryear{da Cunha et al.}{2013a}]{dacunha13} da Cunha E., Walter F., Decarli R., Bertoldi F., Carilli C., Daddi E., Elbaz D., Ivison R., et al., 2013a, ApJ, 765, 9
\bibitem[\protect\citeauthoryear{da Cunha et al.}{2013b}]{dacunha13b} da Cunha E., Groves B., Walter F., Decarli R., Wei\ss{} A., Bertoldi F., Carilli C., Daddi E., Elbaz D., Ivison R., et al., 2013b, ApJ, 766, 13
\bibitem[\protect\citeauthoryear{Daddi et al.}{2004}]{daddi04} Daddi E., Cimatti A., Renzini A., Fontana A., Mignoli M., Pozzetti L., Tozzi P., Zamorani G., 2004, ApJ, 617, 746
\bibitem[\protect\citeauthoryear{Daddi et al.}{2008}]{daddi08} Daddi E., Dannerbauer H., Elbaz D., Dickinson M., Morrison G., Stern D., Ravindranath S., 2008, ApJ, 673, L21
\bibitem[\protect\citeauthoryear{Daddi et al.}{2009}]{daddi09} Daddi E., Dannerbauer H., Stern D., Dickinson M., Morrison G., Elbaz D., Giavalisco M., Mancini C., et al. 2009, ApJ, 694, 1517
\bibitem[\protect\citeauthoryear{Daddi et al.}{2010}]{daddi10a} Daddi E., Bournaud F., Walter F., Dannerbauer H., Carilli C.L., Dickinson M., Elbaz D., Morrison G.E., et al., 2010, ApJ, 713, 686
\bibitem[\protect\citeauthoryear{Dannerbauer et al.}{2009}]{dannerbauer09} Dannerbauer H., Daddi E., Riechers D.A., Walter F., Carilli C.L., Dickinson M., Elbaz D., Morrison G.E., 2009, ApJ, 698, L178
\bibitem[\protect\citeauthoryear{Decarli et al.}{2013}]{decarli13_aless} Decarli R., et al. 2013, submitted to ApJ
\bibitem[\protect\citeauthoryear{Dickinson et al.}{2003}]{dickinson03} Dickinson M., Papovich C., Ferguson H.C., Budav\'{a}ri T., 2003, ApJ, 587, 25
\bibitem[\protect\citeauthoryear{Elbaz et al.}{2011}]{elbaz11} Elbaz D., Dickinson M., Hwang H.S., D\'{i}az-Santos T., Magdis G., Magnelli B., Le Borgne D., Galliano F., et al., 2011, A\&A, 533, 119
\bibitem[\protect\citeauthoryear{Fern\'{a}ndez-Soto et al.}{1999}]{soto99} Fern\'{a}ndez-Soto A., Lanzetta K.M., Yahil A., 1999, ApJ, 513, 34
\bibitem[\protect\citeauthoryear{Genzel et al.}{2010}]{genzel10} Genzel R., Tacconi L.J., Gracia-Carpio J., Sternberg A., Cooper M.C., Shapiro K., Bolatto A., Bouch\'{e} N., et al., 2010, MNRAS, 407, 2091
\bibitem[\protect\citeauthoryear{Grogin et al.}{2011}]{grogin11} Grogin N.A., Kocevski D.D., Faber S.M., Ferguson H.C., Koekemoer A.M., Riess A.G., Acquaviva V., Alexander D.M., et al. 2011, ApJS, 197, 35
\bibitem[\protect\citeauthoryear{Hughes et al.}{1998}]{hughes98} Hughes D.H., Serjeant S., Dunlop J., Rowan-Robinson M., Blain A., Mann R.G., Ivison R., Peacock J., et al., 1998, Nature, 394, 241
\bibitem[\protect\citeauthoryear{Karim et al.}{2011}]{karim11} Karim A., Schinnerer E., Mart\'{\i}nez-Sansigre A., Sargent M.T., van der Wel A., Rix H.-W., Ilbert O., Smol\v{c}i\'{c} V., et al., 2011, ApJ, 730, 61
\bibitem[\protect\citeauthoryear{Kennicutt}{1998}]{kennicutt98} Kennicutt R.C., 1998, ApJ, 498, 541
\bibitem[\protect\citeauthoryear{Koekemoer et al.}{2011}]{koekemoer11} Koekemoer A.M., Faber S.M., Ferguson H.C., Grogin N.A., Kocevski D.D., Koo D.C., Lai K., Lotz J.M., et al. 2011, ApJS, 197, 36
\bibitem[\protect\citeauthoryear{Reddy et al.}{2008}]{reddy08} Reddy N.A., Steidel C.C., Pettini M., Adelberger K.L., Shapley A.E., Erb D.K., Dickinson M., 2008, ApJS, 175, 48
\bibitem[\protect\citeauthoryear{Riechers et al.}{2013}]{riechers13} Riechers D.A., Bradford C.M., Clements D.L., Dowell C.D., P\'{e}rez-Fournon I., Ivison R.J., Bridge C., Conley A., et al. 2013 Nature, 496, 329 
\bibitem[\protect\citeauthoryear{Rosolowski \& Leroy}{2006}]{rosolowski06} Rosolowski E. \& Leroy A., 2006, PASP, 118, 590
\bibitem[\protect\citeauthoryear{Rujopakarn et al.}{2010}]{rujopakarn10} Rujopakarn W., Eisenstein D.J., Rieke G.H., Papovich C., Cool R.J., Moustakas J., Jannuzi B.T., Kochanek C.S., et al. 2010, ApJ, 718, 1171
\bibitem[\protect\citeauthoryear{Sargent et al.}{2012}]{sargent12} Sargent M.T., B\'{e}thermin M., Daddi E., Elbaz D., 2012, ApJ, 747, L31
\bibitem[\protect\citeauthoryear{Sargent et al.}{2013}]{sargent13} Sargent M.T., Daddi E., B\'{e}thermin M., Aussel H., Magdis G., Hwang H.S., Juneau S., Elbaz D., da Cunha E., 2013 (arXiv:1303.4392)
\bibitem[\protect\citeauthoryear{Silva et al.}{1998}]{silva98} Silva L., Granato G.L., Bressan A., Danese L., 1998, ApJ, 509, 103
\bibitem[\protect\citeauthoryear{Walter et al.}{2012}]{walter12} Walter F., Decarli R., Carilli C., Bertoldi F., Cox P., da Cunha E., Daddi E., Dickinson M., et al., 2012, Nature, 486, 233
\bibitem[\protect\citeauthoryear{Wei\ss{} et al.}{2007}]{weiss07} Wei\ss{} A., Downes D., Walter F., Henkel C., 2007, ASPC, 375, 25
\bibitem[\protect\citeauthoryear{Wei\ss{} et al.}{2013}]{weiss13} Wei\ss{} A.,  	De Breuck C., Marrone D.P., Vieira J.D., Aguirre J.E., Aird K.A., Aravena M., Ashby M.L.N., et al., 2013, ApJ, 767, 88
\bibitem[\protect\citeauthoryear{Williams et al.}{1996}]{williams96} Williams R.E., Blacker B., Dickinson M., Dixon W.V.D., Ferguson H.C., Fruchter A.S., Giavalisco M., Gilliland R.L., et al., 1996, AJ, 112, 1335
\end{thebibliography}
\end{document}